\documentclass[12pt]{article}
\usepackage{graphicx}
\usepackage{amsmath}
 \usepackage{longtable}
 \usepackage{color}
\relax
\usepackage{amssymb}

\newcommand{\bo}{\overline{o}}

\def\bo{{\raise.15ex\hbox{\large$\Box$}}}               

\def\face{{\raise.2ex\hbox{$\displaystyle \bigodot$}\mskip-2.2mu \llap {$\ddot
        \smile$}}}                                      

\def\leftrightarrowfill{$\mathsurround=0pt \mathord\leftarrow \mkern-6mu
        \cleaders\hbox{$\mkern-2mu \mathord- \mkern-2mu$}\hfill
        \mkern-6mu \mathord\rightarrow$}       
\def\dvec#1{\vbox{\ialign{##\crcr
        \leftrightarrowfill\crcr\noalign{\kern-1pt\nointerlineskip}
        $\hfil\displaystyle{#1}\hfil$\crcr}}}           

\def\beq{\begin{equation}}
\def\eeq{\end{equation}}

\def\beqx{\begin{displaymath}}
\def\eeqx{\end{displaymath}}

\def\beql{\begin{eqnarray}}
\def\eeql{\end{eqnarray}}

\newcommand{\bea}{\begin{eqnarray}}
\newcommand{\eea}{\end{eqnarray}}

\def\[{\left [}
\def\]{\right ]}
\def\({\left (}
\def\){\right )}

\def\+{\oplus}

\begin{document}

\hbox{\hskip 12cm NIKHEF/2013-038  \hfil}
\hbox{\hskip 12cm IFF-FM-2013/03  \hfil}

\vskip .5in

\begin{center}
{\Large \bf GUTs without guts }

\vspace*{.4in}
{B. Gato-Rivera}$^{a,b}$
\ {and A.N. Schellekens}$^{a,b,c}$
\\
\vskip .2in

${ }^a$ {\em NIKHEF Theory Group, Science Park 105, \\
1098 XG Amsterdam, The Netherlands} \\

\vskip .2in

${ }^b$ {\em Instituto de F\'\i sica Fundamental, IFF-CSIC, \\
Serrano 123, Madrid 28006, Spain} \\

\vskip .2in

${ }^c$ {\em IMAPP, Radboud Universiteit,  Nijmegen}

\end{center}

\begin{center}
\vspace*{0.3in}
{\bf Abstract}
\end{center}

The structure of a Standard Model family is derived in a class of brane models with a $U(M)\times U(N)$ factor, 
from two mildly anthropic
requirements: a massless photon and a universe that does not turn into a plasma of massless charged particles.
If we choose $M=3$ and $N=2$, the only option is shown to be the Standard Model with an undetermined number of families. We do not
{\it assume} the $U(1)$ embedding, charge quantization, family repetition, nor the fermion representations;  all of these features are {\it derived}, 
assuming a doublet Higgs. With a slightly stronger assumption even the Higgs representation is determined.
We also consider a more general class, requiring an asymptotically free strong $SU(M)$  (with $M \geq 3)$ interaction 
from the first factor and an electromagnetic $U(1)$ embedded in both factors. We allow Higgs symmetry breaking of the $U(N)\times U(1)$ flavor group
by at most one Higgs boson in any representation, combined with any allowed chiral symmetry breaking by $SU(M)$. 
For $M=3$ there is a large number of solutions with an unbroken $U(1)$. In all of these, ``quarks" have third-integral
charges and color singlets have integer charges in comparison to leptons. Hence Standard Model charge quantization holds for any $N$. 
Only for $N=2$ these models allow an $SU(5)$ GUT extension, but this extension offers no advantages whatsoever for understanding
the Standard Model; it only causes complications, such as the doublet-triplet splitting problem. 
Although all these models have a massless photon,   all except the Standard Model are ruled
out by the second anthropic requirement. In this class of brane models the Standard Model  is realized as a GUT with its intestines removed, to
keep only the good parts: a GUT without guts.

\vskip .5truecm

\noindent

\newpage

\section{Introduction}

Susy-GUTs are the pinnacle of the symmetry-based approach to particle physics, which has brought us
the phenomenal success of the Standard Model. From the perspective of symmetry it was, and still is, hard to
believe how this could be wrong. And yet susy-GUTs have led to an impressive series of false predictions and expectations.
Around 1983 the confirmation of $SU(5)$ GUTs \cite{Georgi:1974sy} (without supersymmetry) seemed just around the corner, but the expected proton
decay was not found. A decade later, improved precision from LEP data made it  clear that the three gauge couplings did not pass
through a single point as was once believed, but that they  do so if supersymmetry partners were included in the running \cite{Amaldi:1991cn}. This
introduces a new parameter, the supersymmetry breaking scale $M_S$, and hence in this scenario the three gauge couplings are fitted
using three parameters, $M_S, M_{\rm GUT}$ and $\alpha_{\rm GUT}$, instead of the two of minimal, non-supersymmetry $SU(5)$ GUTs.  
The only reason why the coupling convergence may nevertheless be
called miraculous is that this new parameter was found to coincide with the weak scale, as required for supersymmetry  to solve the hierarchy problem.

However, this too has now run into trouble because of the first round of LHC results. The observed Higgs mass value, the absence of
direct susy signals and the impressive results on flavor physics from LHCb have put the idea of low energy supersymmetry under severe
stress, and cast doubts on the relevance of the aforementioned miracle. It is true that if supersymmetry is still found during the next run
of LHC, starting in 2015, the convergence of the three coupling constants will still be acceptable. The dependence on the supersymmetry breaking
scale is only logarithmic, and furthermore there are intrinsic uncertainties due to unknown (and in general unknowable) threshold corrections at
the GUT scale. But the two or three orders of fine-tuning of the weak scale that would already be needed make the argument less convincing.

GUTs were once believed to follow naturally from string theory. But this is probably just a ``lamppost effect". The most easily accessible compactifications
are the symmetric ones, for example the Calabi-Yau compactifications of the $E_8\times E_8$ heterotic string, leading to $E_6$ unification. But under more
general conditions many other gauge groups are possible, and it is not at all clear that GUT models are favored in the full heterotic landscape. Similar comments
apply to other corners of the landscape, such as brane models. We get GUTs  because we want them, 
and not because string theory requires them. 
Indeed, string theory casts some doubts on the ultimate validity of the traditional symmetry approach that has dominated particle physics for decades.
Although internally string theory is strongly symmetry-based,
its solutions, forming the string theory landscape, are more anarchic than symmetric. This is true for the choice of parameters as well as 
for the gauge groups and representations. We will focus on the latter in this paper.

%
Has the time arrived to throw the entire susy-GUT idea into the dustbin of history? This may seem an appalling idea, because 
GUTs provide two important insights in the Standard Model: gauge coupling
unification and a determination of the group structure of the Standard Model families, including the correct quantization of electric charge. 
But perhaps the former can be dismissed as a coincidence at the few percent level. If the 
$SU(3)$, $SU(2)$ and $U(1)$ couplings are perturbative at the Planck scale, they will evolve roughly in this manner. The $SU(3)$ coupling increases towards the
infrared until it reaches a Landau pole, the $U(1)$ coupling decreases and the $SU(2)$ coupling lies somewhere in between. This already generates the illusion of 
approximate convergence. The susy-GUT convergence miracle is not fully explained by this, but in 1983 many were equally convinced by an approximate non-supersymmetric convergence
that  turned out to be wrong. And furthermore the Standard Model harbors a much more impressive coincidence, 
the Koide mass formula\footnote{This formula is $m_e+m_{\mu}+m_{\tau}=\frac23 (\sqrt{m_e}+\sqrt{m_{\mu}}+\sqrt{m_{\tau}})^2$, where the coefficient $\frac23$ is exactly in the
middle between its mathematical  bounds, $\frac13$ and $1$. Note that this formula relates pole masses at different mass scales, so it is hard to reconcile it with the renormalization group.} 
 \cite{Koide:1983qe}, that most people 
regard as indeed  a mere coincidence. 

Charge quantization is harder to dismiss. In  its currently known form the Standard Model gauge group allows particles with charges that are arbitrary real numbers. 
Anomaly cancellation imposes some restrictions, but  one can add
non-chiral matter with arbitrary charges. In fact, one can even add irrationally charged {\it chiral} matter that gets a mass from the Standard Model Higgs boson (see section \ref{HiggsMultiplets}). 

Just like the smallness of $\bar\theta_{QCD}$,\footnote{By $\bar\theta_{QCD}$ we mean here the observable
quantity that contains a contribution from quark mass diagonalization.} 
charge quantization is a problem that demands a solution, because  they
would otherwise be weird coincidences that are not even anthropically explained. 
The comparison to $\bar\theta_{QCD}$ is appropriate for
{\it absolute} charge quantization, by which we mean the quantization of confined as well as unconfined
electric charge in (suitable normalized) integer  units. 
Theoretically, the group $SU(3)\times SU(2) \times U(1)$ allows any deviation from those integers, just as $\bar\theta_{QCD}$ is allowed to be any angle.
Experimentally, the observed charge ratios are extremely close to mathematically special values, just as $\bar\theta_{QCD}$. 
Nothing in the theory excludes future observations that violate these facts. The limits on non-integer charges are extremely good (see \cite{Perl:2009zz}), but
they are just limits. Viewed in this way, absolute  charge quantization is an even more serious problem than the strong CP problem, and an excellent indicator for physics beyond
the Standard Model.
The extreme possibility of ending up with
just the current Standard Model, coupled to a theory of gravity that has nothing to say about charge quantization is extremely unpalatable and implausible. 

Fortunately, if the theory of gravity is string theory, at least the absolute quantization problem is solved.
The duality structure of string theory probably implies quantization of $U(1)$ charges in some rational units, and it certainly 
does in all explicitly known constructions. However, this does not  guarantee that the correct  correlation between color and fractional charge
observed in the Standard Model comes out. 
In field theory, GUTs solve {\it both} problems (absolute quantization and charge integrality of color singlets) in a beautiful way. But given that string theory
already solves the first problem, 
do we really need both string theory {\it and} GUTs?

String theory  adds two important ingredients to the understanding
of the Standard Model: small representations and anomaly cancellation. There is no restriction on the size of representations 
in field theory. Large representations may ruin asymptotic freedom, but that is not a fundamental consistency requirement. Anomaly
cancellation is an ad-hoc and {\it a posteriori} consistency requirement in field theory, but follows inescapably from string theory.   
Anomaly cancellation in $SO(10)$ spinor representations is an interesting exception. This case provides the strongest argument against
the point of view we will present here. But this would be  a strong argument only in alternatives to string theory that unambiguously
predict $SO(10)$ (we are not aware of any, but noncommutative geometry once seemed to hint in that direction  \cite{Chamseddine:1993is}).
If one considers $SO(10)$ in the context of string theory, it is just one of many options, and much of the beauty is lost \cite{Lebedev:2006kn}. Furthermore, since the anomaly cancellation follows from string theory anyway, also the main QFT argument in its favor is nullified.


The restrictions on representations string theory offers are a huge step towards understanding the particle content of  the Standard Model.
However, it has been difficult to get the details right in string theory embeddings of the Standard Model. In heterotic strings, even starting from GUTs,
the spectrum always contains fractionally charged color singlets  \cite{Schellekens:1989qb} (see also \cite{Wen:1985qj,Athanasiu:1988uj}), if one uses the
standard $SU(5)$-related $U(1)$ normalization.
These fractionally charged particles may be massive and unobservable.
Examples where all fractionally charged particles have Planck masses can indeed be found \cite{Assel:2009xa,Blaszczyk:2009}, 
but these examples appear to occur only rarely \cite{Assel:2010wj,GatoRivera:2010gv,GatoRivera:2010xn}. This is phenomenologically acceptable, but not
an explanation for their absence, and also a rather disappointing outcome from the GUT point of view.
 In brane models, charge quantization is automatic in one class of models (those with $x=0$ in the classification of
\cite{Anastasopoulos:2006da}) and semi-automatic in  another. But these models are constructed in a bottom-up way  with the requirement that all known types of particles
can be obtained with their correct charges. That is not an explanation either. Furthermore the set of allowed representations includes rank-2 tensors and bi-fundamentals
that do not appear in the Standard Model spectrum. The condition of tadpole or anomaly cancellation allows many other solutions than just the Standard Model. These are
eliminated by hand, by discarding spectra that do not match the data. The most successful top-down description of the Standard Model 
structure is provided by 
brane GUTs and F-theory GUTs.
However,  there is no argument why  $SU(5)$ would be
preferred in the string landscape.   In brane models, one could just as easily start with separate $U(3)$ and $U(2)$ stacks. Furthermore the first 
explorations of $SU(5)$ brane GUTs \cite{Cvetic:2002pj} found an overwhelming abundance of spectra with symmetric tensors, the $({\bf 15})$ of $SU(5)$, 
which are  rejected on phenomenological grounds and not for any fundamental reason.

This is an unsatisfactory state of affairs if one adopts the old paradigm of uniqueness and symmetry. Embedding GUTs in string theory should have improved our understanding, but just the opposite happens. It is also an unsatisfactory state of affairs from the opposite point of view that our
Universe is merely one of many options in a multiverse. Charge quantization, the family structure and family repetition demand a better explanation than just  chance.

In this paper we will  show how one can essentially derive the Standard Model from very few assumptions. The assumptions include a certain string realization, some
anthropic requirements and some conjectures about landscape statistics, which only play a secondary r\^ole. 

This paper is organized  as follows. In the next section we present our assumptions and how they are motivated. In recent years we often heard the Higgs advertised 
as the origin of all mass. Although this is incorrect for several reasons, it is true that it gives a mass to all charged quarks and leptons, and turns a fully chiral theory into a fully non-chiral one.
Roughly speaking, our main point is understanding why this can be viewed as an anthropic requirement, and finding all solutions to this condition within a class of theories. 

Section \ref{HiggsMultiplets} is an intermezzo on
the Higgs mechanism. We investigate how a Higgs system must be built to make sure that all components of the broken fermion multiplets acquire masses from allowed
Yukawa interactions. This leads to a surprisingly non-trivial structure, and for the breaking $SU(N)\times U(1) \rightarrow SU(\!N\!-\!1\!)\times U(1)$ by a Higgs in the vector representation of $SU(N)$,
the Higgs system of the Standard Model turns out to be the simplest possibility, for any fermion representation and any value of $N$. 

However, this section can be skipped 
if one only wants to understand the main result, which is presented in section \ref{Derivation}. We present it as a path towards a derivation of the Standard Model, because it 
almost looks as if it comes out inescapably from some simple algebra, just by requiring complexity to emerge from the simplest 
 ingredients. The ingredients are two unitary  brane stacks (denoted $U(M)\times U(N)$) with some additional neutral branes that do not contribute to the gauge group.
We find that  the
Standard Model  can be characterized by objective criteria that define it uniquely within this part of the landscape.

The discussion in section \ref{Derivation} is limited to $U(M)\times U(N)$ brane models with $M \geq 3$.
In general, the discussion is hampered by poorly understood
strong interaction dynamics, since there are two non-abelian groups competing with each other. Therefore
we limit ourselves to the best understood subclass:  we assume that there is one strong interaction gauge group that dominates, and  that this group is the
$SU(M)$ subgroup of the first factor. In particular it must be asymptotically free. 
Furthermore we require that there is an electromagnetic $U(1)_{\rm em}$ that must be embedded in the $U(1)$ factor of the first brane
and in the $U(N)$ group of the second one. The $U(N)$ group must be broken, and we assume that this is done by a single Higgs boson, for reasons explained in the next section. Furthermore it may
be broken by a chiral symmetry breaking mechanism of the strong interaction gauge group $SU(M)$. We consider any dynamical breaking that leads to a non-chiral theory, so that all fermions can get 
a mass. Indeed, we even allow chiral symmetry breaking alone to produce the non-chiral theory without the help of the Higgs mechanism, as it would in the Standard Model if the Higgs were absent.
Our results are most reliable for $SU(M)$ theories with only vectors and anti-vectors of $SU(M)$, {\it i.e.} no chiral tensors. 
If $M=3$ and $N=2$ we can
show that these tensors must be absent using an argument in the lepton sector. In this class we have maximal control over the dynamics. 
For $M=3$, and $N\not=2$ a similar argument can be made, but it requires a slightly stronger assumption. For $M > 3$ a  still stronger assumption is necessary. 

In section \ref{Exceptional} we discuss some cases not covered by the assumptions of section \ref{Derivation}. This includes the possibility that the electromagnetic $U(1)$ is
just the $U(1)$ of the $U(M)$ group or of the $U(N)$ group. This yields some solutions to our anthropic conditions, but these spectra have no leptons at all. It may also happen that the Higgs breaks the high energy $U(1)$ factor completely, so that a replacement has to be
generated dynamically at low energy.
Furthermore we find some solutions with $M < 3$, which include spectra without any strong interactions, or with $SU(2)$ or $SO(4)$ strong interactions in the low energy
gauge group. We also discover some interesting relatives of the Standard Model. None of the models in section \ref{Exceptional} look anthropically viable, but our aim is to use only the
least questionable anthropic assumptions, and hence we simply keep these theories on our list of acceptable models. 

The phenomenology of this class of brane models depends on their concrete realization, which is not the main focus of this paper. 
Nevertheless, 
in section \ref{GutlessGut} we comment briefly on a few  phenomenological features of the brane realization of the Standard Model we end up with. 
It looks like an $SU(5)$ GUT with all of its non-essential features  removed: a GUT without guts\rlap.\footnote{It has
been suggested to us that ``guts without GUTs" might also have been a suitable title.} 
 
However, our main goal is to demonstrate that the structure of the Standard Model can be understood without GUTs. We were pleasantly
surprised to discover that not only it can be understood, but that in addition by considering alternatives we were able to  appreciate its remarkable structure far better than ever. 


\subsection{Terminology}

In this paper we discuss alternatives to the Standard Model that might exist in other universes. To prevent confusion we should
clarify our use of the particle physics terminology of our own universe.
In most of the discussion there will be one  gauge group that is asymptotically free and assumed to be the strongest at low energies.
All fundamental fermions coupling to this gauge group will be called ``quarks", and all strong interaction bound states will be
called ``hadrons". Since we do not want to make any assumptions about nuclear physics, what we actually mean by hadrons is {\it any}
bound state that is due to the strong interaction gauge group, including the nuclei in our own universe.
We do not know a general, objective distinction between these two kinds of bound states ({\it i.e.} nuclei and nucleons). 
Any fundamental fermions not coupled to the strong gauge group will be called   ``leptons".
There may exist other non-abelian gauge groups, even after Higgs symmetry breaking. These may form bound states of leptons with each other 
or with hadrons. Leptons that do not couple to any non-abelian gauge groups will be called ``free leptons". 

\section{Assumptions and Motivations}

\subsection{String Theory}

The main r\^ole of string theory in our argument is to limit the choice of representations, to provide a rationale for anomaly cancellation, and 
to  give rise to absolute charge quantization (the latter will be explained more precisely below). 
Our arguments are not limited to string theory, and would work in any theory that provides these ingredients, but something beyond QFT is needed.
The  unlimited
set of possibilities of quantum field theory would make the problem totally intractable. String theory limits the choice of representations in any region of the landscape.
They are restricted to fields with conformal weight $h \leq 1$ in heterotic strings, and to rank two tensors and bi-fundamentals in open string/intersecting brane models.
 In this
paper we will consider intersecting brane models. We will not focus on any particular realization of such models (see \cite{Blumenhagen:2005mu} for an overview of many possibilities).
The restriction brane models impose on massless representations is robust and  does not depend on particular realizations. 
Nevertheless, this restriction on allowed representations is still our strongest assumption. 

{In this paper we limit ourselves to the case where at most two unitary brane stacks
provide the low energy gauge group. Note that this is not the usual way of counting brane stacks, since we do allow an arbitrary number of branes that do not contribute to the low-energy gauge group.
The restriction to two unitary stacks is made for two reasons. First of all, it is the smallest number of stacks for which a solution to our anthropic conditions can be found. Secondly, we already know that
this class contains $SU(5)$ GUTs, and one of our main goals is to investigate the importance of $SU(5)$ unification. Other brane configurations  that have been discussed in the literature as Standard Model candidates,
such as Pati-Salam models \cite{Pati:1974yy}, flipped $SU(5)$ \cite{Barr:1981qv}, the Madrid model \cite{Ibanez:2001nd} and trinification \cite{Glashow:1984gc} require more stacks participating in the low energy gauge group. 
See \cite{Anastasopoulos:2006da} for a systematic survey. We hope to extend our analysis to these cases in the future.

Examples of other theories offering restrictions on field theory content where our approach can be tried are heterotic strings and F-theory. The latter allows matter representations in addition to those of brane models, which would
be considered non-perturbative from the brane model point of view. See \cite{Grassi:2013kha} and references therein. Outside string theory, one may consider noncommutative geometry \cite{Chamseddine:2007hz} or
noncommutative gauge theory \cite{Chaichian:2001mu,Chaichian:2001py} (not to be confused with each other). These papers are indeed aimed at understanding the standard model structure, but they
do not consider anthropic conditions.}

{Apart from a restriction on representations, }string theory will also provide
justification for anomaly cancellation. In a few cases this includes ``stringy" anomaly cancellations that  do not exist in field theory. There are
$SU(2)$ and ``$SU(1)$" cubic anomaly relations that follow from tadpole cancellation in brane models, and that look like an extrapolation of the
field theoretic expressions for $SU(N)$ \cite{Ibanez:2001nd,Dijkstra:2004cc}. These $SU(2)$ anomalies were recently re-emphasized in \cite{Halverson:2013ska}. 
However, we will find that in the case of most interest the stringy  $SU(2)$ relation is redundant, and the conventional field-theoretic relations suffice.

\subsection{Anthropic assumptions}

The basic idea of an anthropic assumption is to  try and argue that all the alternatives to the Standard Model that some fundamental theory may offer do not  allow the
existence of observers, and hence cannot be observed. See \cite{Schellekens:2013bpa} for a recent review and further references.
One can easily go too far with this, and make the argument hinge on fine details like Carbon abundance.
But our  anthropic assumptions are of an extremely mild kind. 
The main  assumption
is that the low energy theory contains at least one abelian factor coupled to a set of long-lived massive charged particles.  
Electrodynamics in all its manifestations plays such a crucial r\^ole in our own Universe that we cannot even begin to imagine life without it. Giving a mass to the photon
has many catastrophic consequences for our kind of life. This motivates the assumption  that QED will be essential for life in general, or at least that the life without it is so severely challenged that the vast majority of observers in the multiverse will find themselves in
an electromagnetic universe.
We do not
put strong constraints on the spectrum of charges, except that something too simple will not do. Experience from our own Universe shows that just hydrogen and helium will not give rise to enough complexity. 
We are also not assuming that the charged building blocks can be divided into hadrons or leptons,
nor that they are elementary or composite, nor that they have any particular mass ratio, nor that they have any special choice of charges. In our Universe charges like
$-1$, $1,6, 8$ and  a few others are needed for life, but we will not impose this.

We {\it do} assume that these building blocks of life have masses well below the Planck scale. 
Otherwise
complex structures would be crushed by gravity. If a complex entity consist of $N$ building blocks, the masses of these blocks must be less than $N^{-1/3} M_{\rm Planck}$ to
prevent this. This is just the usual argument that  the maximal amount of protons in a star  is about $10^{57}$, applied to  the intelligent beings themselves. For a
a human brain this simple argument would already require a hierarchy of at least $10^{-9}$, for the average mass ratio of 
building blocks (like protons or nuclei in our Universe) and the Planck scale. This explains half the hierarchy on a logarithmic scale, but
 then we only have just enough of a hierarchy to prevent collapse of brains into black holes. Clearly an even larger hierarchy is needed,
and there are indeed anthropic arguments that get much closer to the observed hierarchy, 
but they make much stronger assumptions about the laws of physics. See \cite{Agrawal:1997gf} and \cite{Schellekens:2013bpa} for further
discussion and references.

Nothing we assumed so far precludes using pure QED with elementary particles 
as building blocks for life. This  idea encounters numerous
challenges. There would be no fusion-fueled stars, but degenerate stars, like neutron stars or white dwarfs in our Universe,  could take over their r\^ole \cite{Adams:2008ad}).
There would be  no possibility for Big Bang or stellar nucleosynthesis. Instead one needs a mechanism analogous to baryogenenis in our Universe, where a net surplus 
of fundamental particles over anti-particles is created for all relevant building blocks of matter. It is totally unclear how to realize that in pure QED.
But we will focus here on another problem, namely the huge hierarchy
 problem caused by a substantial number of light  particles\rlap.\footnote{The particles must be light, but not massless, since massless charged particles have
 disastrous implications. We will return to this later.}

\subsection{The Gauge Hierarchy}

In the Standard Model the proliferation of light particles
is solved by obtaining all masses of the light fundamental fermions from a single scale, the mass of the Higgs boson. 
Note that the strong scale (set in a natural way by means of dimensional transmutation)
dominates the proton
and neutron masses, but that is only true because the quark masses are small. 


Inspired by this we add the Higgs mechanism to our list of assumptions. We will require that the fundamental theory has some high energy gauge group $G$, broken
at some small scale by the vev of a Higgs to a subgroup $H$. This is  not really an anthropic assumption, but an assumption about landscape statistics. We are assuming
that it is statistically less costly to make a single scalar light than a number (at least three) of fermions. If that is the case, one would expect  that the high energy
gauge group has a chiral, massless spectrum, with just one light scalar. Anything non-chiral, and all other bosons would be very massive, because it is
too unlikely for them to be light. Of course this is precisely what we observe.

\subsubsection{Naturalness and the Hierarchy Problem}\label{Naturalness}

The aforementioned statistical assumptions may
seem to run counter to the idea of technical naturalness. The $\mu^2$ mass parameter in the Higgs potential receives quadratic corrections
from any high scale, so that its perturbative expansion takes roughly the following form
\begin{equation}
\label{quad}
\mu^2_{\rm phys}= \mu^2_{\rm bare} + \sum_i a_i \Lambda^2 \ ,
\end{equation}
For simplicity we use here a single large scale $\Lambda$.  The existence of a hierarchy problem is indisputable if there exist particles
with masses larger than the weak scale. In string theory there are particles with Planck masses, and hence in this context one cannot solve the
hierarchy problem by denying its existence. 
Eq. (\ref{quad}) does not imply that $\mu^2_{\rm phys}$ is of order
$\Lambda^2$, but only that in a sufficiently large ensemble of theories with coefficients $a_i$ of order 1, the fraction of theories
with a desired mass scale  $\mu_{\rm phys}=m$ is of order 
$m^2/\Lambda^2$. 

By contrast, technically natural parameters $\lambda$ renormalize as
\begin{equation}
\label{logs}
\lambda_{\rm phys}= \lambda_{\rm bare}\left(\sum_i b_i {\rm log}(\Lambda/Q)\right)
\end{equation}
where $Q$ is some low energy reference scale. The important difference with (\ref{quad}) is not only that the corrections are logarithmic, and hence of order 1, but
also that the corrections are all proportional to the parameter itself. Hence if the parameter is small, it stays small.

However, whereas (\ref{quad}) determines the statistical distribution of the parameter, (\ref{logs}) does not. Any fundamental distribution of $\mu^2_{\rm bare}$ is washed
out by the loop corrections, but this has the advantage that we can at least estimate the statistical cost. This is not the case for (\ref{logs}), since knowing
the distribution requires knowing something about the fundamental theory. If, for example, a non-chiral fermion mass is given by $\lambda v$, where $v$ is a Planck scale
vev (one may think of a modulus), and if $\lambda$ has a flat distribution, the statistical cost of a single light fermion with mass $m$ is $m/\Lambda$, and three light
fermions would be more costly than a single boson, {\it i.e.} $(m/\Lambda)^3 \ll (m/\Lambda)^2$. The observed Yukawa couplings for quarks and leptons do not suggest a flat, but scale invariant distribution
\cite{Donoghue:1997rn}. However, such  a distribution requires a cut-off at small $\lambda$, or else  exponentially small values are highly preferred. This is apparently
not the case for the observed Yukawa couplings, nor for masses of vector-like fermions (since we have not seen any yet). In some string theory examples, those couplings
originate from exponents of actions, which are given by the surface area of a world-sheet enclosed by three branes (word-sheet instantons). On a compact surface, these
areas are geometrically limited. This would lead to a sharp fall-off of the distribution at small values of the coupling constant, which could well be much stronger than a 
power law. 

All of this shows that the intuitive idea that ``technically natural" always wins against ``technically not natural" is not a foregone conclusion. For {\it technically
not natural} parameters the statistical cost can be computed assuming all terms in (\ref{quad}) have their natural size. But for {\it technically natural} parameters we need
to know the unknown cost of a parameter being very far from its natural value. We are assuming that for three or more fermions the latter is higher.
Then statistically a single Higgs always wins against three or more non-chiral light fermions. Basically we are viewing the Higgs mechanism as a {\it solution} rather than the {\it cause} of the hierarchy problem!
An additional advantage of this assumption is that it is very unlikely that there is more than one Higgs. Models requiring several low scale Higgs mechanisms to
arrive at atomic physics  are severely challenged statistically in comparison to the Universe we observe.   

The previous argument would be  more convincing if the Higgs hierarchy problem is  reduced by low energy supersymmetry (or other mechanisms such as large extra dimensions or compositeness). 
Then it would be much more
plausible that  the statistical cost
of a single Higgs scale outweighs that of three or more fermions \cite{Gedalia:2010iy}. 
However, since the general spirit of this paper is to see how far one can get with the minimal
amount of symmetry, we will not assume low energy  supersymmetry. It is  not clear whether the cost of low energy supersymmetry (combined with the cost
of avoiding anthropic disasters such as dimension four proton decay) outweighs the benefits 
\cite{Douglas:2004qg,Susskind:2004uv,Denef:2004cf,Douglas:2006es,Douglas:2012bu}. The absence of any signs of supersymmetry during two decades of searches
provides circumstantial evidence against that. Nevertheless, most of our results remain valid in the presence of low energy supersymmetry, but to keep the discussion as simple
as possible we will not take it into account. Of course, the argument in favor of just a single Higgs is weakened if there is low energy supersymmetry\rlap.\footnote{In the MSSM there are
two Higgs fields, but since they align to break the symmetry in the same way, this is a single Higgs system from our point of view.}
Other ideas for solving the
Higgs hierarchy problem that involve new gauge groups and fermions (technicolor and related ideas) cannot  be taken into account as easily, because they imply
drastic changes in the gauge group and representations. As we will see, they enter the discussion in a different way.

\subsection{The R\^ole of the Higgs Mechanism}

The Higgs mechanism will play an essential r\^ole in our argument. The fact that the Higgs gives mass to all quarks
and leptons has been extensively advertised, especially after the discovery of the Higgs boson. In the next subsection we will discuss why that is important, and how it
can be turned into an argument that selects the Standard Model.  We illustrate this by considering the
Standard Model without a Higgs boson. 

\subsubsection{Lessons from the Higgsless Standard Model}

The discussion in Sec. (\ref{Naturalness}) suggests an important question:
why do we need a Higgs mechanism at all? If having more Higgses is  dramatically less probable, then having no Higgs at all would be highly preferred statistically, provided the light fermion masses are generated dynamically.
Without a Higgs mechanism, all chiral fermions would remain chiral, and hence massless. 
In the Standard Model, it is known that  in that case chiral symmetry
breaking in QCD does part of the job of the Higgs (see \cite{Quigg:2009xr} for an extensive discussion). 
The W and Z acquire a mass by eating the three pions. The photon remains massless, and there would be
massive baryons, with masses entirely due to QCD. If we only consider one family  (to avoid 
degeneracies in the baryon spectrum) we would get 
protons and neutrons. The mass splitting between them is only due to electromagnetism and the weak interactions, 
and this would most likely make the neutron the lighter of the two. This fact, and
even more the absence of the pion implies  big changes in nuclear physics, and probably by itself this has already fatal consequences. But we do not want to rely on arguments like these. 
This assumes too much about
the specifics of the Standard Model to serve as a useful general criterion. 

A much more useful criterion is the presence of massless leptons. Since the Bohr radius of a 
Hydrogen atom would go to infinity and the binding energy to zero, this implies that  they would not form bound states, and there would not be atomic physics. But even then, in alternative universes
there might exist  massive leptons in addition to the massless ones, or there might exist stable massive hadrons 
playing the r\^ole of 
leptons in our Universe. One could even imagine a scenario where  two oppositely charged fermionic hadrons  have a large mass ratio, 
to mimic the electron and proton  in our Universe. But even if that is not the case, it is hard to rule out  the possibility that oppositely
charged hadrons of any mass or spin
might have a sufficiently
interesting electromagnetic bound state spectrum. However, even if the massless charged particles do not participate in atomic physics, their mere existence is likely to have
fatal consequences, since they render the vacuum unstable to charged particle pair creation  \cite{Quigg:2009xr}. Such pairs would be produced at
no cost by photons emitted in any interaction, and furthermore the Thomson  amplitude for scattering of photons with these charged particles becomes infinite in the zero mass limit.
Hence the Universe turns into an opaque plasma of massless charged particles.
This is a useful and very generic reason to forbid such
particles. This is the first lesson we learn from the Higgsless Standard Model.

In the hadronic sector it would not really be a problem if the Higgs does not give mass to quarks, because, as already mentioned, the strong interactions would at least
give the baryons a mass. Furthermore in our Universe one quark, the up quark, {\it is} nearly massless.
But even if quarks do not get a mass, at the very least the coupling of the photon to the quarks must be non-chiral. Otherwise strong interaction chiral
symmetry breaking would make the photon massive, and since we started with the assumption that 
QED is an essential ingredient, clearly that would be unacceptable. This is the fate of the Standard Model $Y$-boson when QCD gets strong: its photon acquires a 
mass.
This is the second lesson. 

The third lesson is that this is not necessarily a fatal problem. Even though $Y$ is broken, there is a linear combination of $Y$ and $T_3$ that is non-chiral with respect to QCD, and
that remains unbroken. This is the usual QED generator. Hence to rule out a theory  it is not sufficient to demonstrate that a candidate photon acquires a mass, but we should also check that there
is no other abelian gauge boson that could take over its r\^ole. This is not how it works in our Universe, where the Higgs selects the photon, but this might be the way the problem
is solved in other universes. 

Another important feature of the Standard Model  Higgs mechanism is that it can give different masses to all quarks and leptons. 
On the other hand, a dynamical Higgs mechanism gives all
quarks a mass related to the fundamental scale of the strong interaction, and does not give mass to leptons. 
Then there is no way to tune these masses to special values. In our Universe, there are good reasons to believe that at least
the $u$, $d$ and electron masses have  special anthropic values (see \cite{Schellekens:2013bpa} and references therein for further
discussion).

\subsection{The Scope of this Approach}

The latter argument might be invalidated in complicated theories with many gauge groups with separately
tunable couplings. Such models have been discussed in attempts to obtain the Standard Model quarks and leptons as composites. It is clear that it is not possible to study
these models in sufficient detail to say anything about the complexity of their spectrum, not to mention intelligent life. 

Since we clearly cannot study every imaginable gauge theory, what is it that we may hope to achieve? What we can do is to study gauge theories that are under sufficient control,
either by being sufficiently similar to the Standard Model or by being sufficiently simple. In this paper, we first focus  on a class of gauge theories with a gauge group 
$SU(3)\times SU(2)\times U(1)$ and a Higgs mechanism, and we  demonstrate that in this class the Standard Model fermion representations
and $U(1)$ choice is the only solution to our conditions. This can be shown using only the best-understood strong interaction theory, namely QCD with only triplets and anti-triplets.

But it is nearly
inevitable to ask what happens if we weaken some of the starting assumptions, by changing the non-abelian factors to $SU(M)\times SU(N) \times U(1)$. We will find that one quickly gets into lesser-known
territory. In addition to quarks and anti-quarks  there may be higher rank tensors. Furthermore, the non-abelian group remaining after the Higgs mechanism may contain two
or more factors. In that case, the spectrum may depend on which of these groups is the strongest at low energies. If they are of comparable strength, the problem becomes
even harder. Although a lot has been learned about strongly interacting gauge theories in the last decades, through methods such as  Seiberg duality \cite{Seiberg:1994pq} or the AdS/CFT correspondence \cite{Maldacena:1997re}, 
there is no general procedure to deal with strongly interacting chiral gauge theories. If we get too deep into this territory, we will have to give up.

So we can put gauge theories 
(with fermions and a Higgs)  into three categories: acceptable, not acceptable and undecidable. 
We call them acceptable if they have a spectrum with
at least one massless abelian vector boson and no massless charged particles, and unacceptable if they clearly do not. But there is going to be
a grey area that is too complicated to analyze, and this area will become larger as we move further away from the Standard Model. An
acceptable theory is by no means guaranteed to allow life. It still has a large number of anthropic hurdles to overcome, some of them very 
severe, such as sufficient stability of the constituents of matter. It is {\it a priori} clear
that the ``acceptable" category defined above will not just contain the Standard Model. For example,
there exists a sequence of theories with color group $SU(M)$ that satisfies all our criteria  and might be just as good, as discussed in section \ref{HiggsMultiplets}. These are the same as the Standard model, but with
$SU(M)$ instead of $SU(3)$, and quark charges that are multiples of $\tfrac{1}{M}$  ($\tfrac{1}{2M}$ if $M$ is even) instead of $\tfrac13$.
For any odd $M$  the quarks in this theory may combine into
charged baryons that could combine in their turn into nuclei. For even $M$ there would only be bosonic bound states, but in both cases we put these theories in the ``acceptable" category.  In the brane models we consider, only the $M=3$ member of this series occurs, but in more general brane models one
can realize other values of $M$. 

Perhaps one day one can find fundamental, anthropic or statistical arguments to determine the structure of the Standard Model completely,
including the number of families, but the more limited scope of identifying the Standard Model as a very special object in its local environment is
already a worthy goal.

\subsection{Caveats}

There is broad agreement on the fact that an $SU(M)$ gauge theory with $F$ flavors of quarks and anti-quarks will break its $U(F)_L\times U(F)_R$ flavor group to a diagonal vector 
subgroup $U(F)_V$, as long as the number of flavors is small enough to respect asymptotic freedom of $SU(M)$. This breaking will 
result in $F^2\!-\!1$ Goldstone bosons (``pions"), plus
one heavier boson. The latter is related to the axial $U(1)$, which is anomalous with respect to $SU(M)$ color. This plays no r\^ole in the following.
Our understanding of chiral symmetry breaking is mainly based on an empirical and a theoretical argument. The empirical argument relies on the fact that the chiral symmetries in the Lagrangian are
clearly not realized in the baryon spectrum of the 
Standard Model, and the only way to understand this is that the vacuum is not invariant under the symmetry. Strictly speaking, this empirical argument holds for six quarks, with definite
masses, coupling to a photon and to weak interactions. Furthermore, these masses and interactions are not irrelevant for the final outcome. The standard quark condensate takes
the form $\overline{u}u+\overline{d}d+\ldots$, where the left components are from the weak doublets and the right ones from the singlets. Such a condensate transforms in a representation
$(F,F)$ of the flavor group $U(F)_L\times U(F)_R$. The direction the condensate chooses is clearly influenced by the quark masses and/or the electromagnetic interaction. Without that,
one could use $U(F)_L$ rotations to rotate the condensate from $\sum_i \bar q_i q_i$ to $\sum_i \bar q_i U_{ij} q_j$ where $U_{ij}$ is a unitary matrix. For $U \not= {\bf 1}$ this condensate
is not invariant under electromagnetism, and would break it, giving a mass to the photon. The reason that this does not happen must be that tiny electromagnetic effects and/or the quark masses generate a potential in
the otherwise flat $U(F)\times U(F)$ space, so that a preferred direction exists.

We have to worry about this fact if we want to claim that a chirally embedded $U(1)$ is always dynamically broken. Could it be that the very presence of this $U(1)$ somehow influences the outcome?
Indeed, a fundamental scalar field  in the representation $(F,F)$ can break $SU(F)\times SU(F)$ to a diagonal subgroup, but also to $SU(F\!-\!1)\times SU(F\!-\!1)$ \cite{Li:1973mq}, depending on the
relative size of terms in the Higgs potential.
So perhaps it is possible that only neutral pairs condense. The problem with this possibility is that we would still be left with chiral, electrically unpaired quarks, that somehow have to find a way
to end up in a bound state. A left-handed fermionic bound state would not have a matching right-handed partner, so cannot be massive.
One possibility is that the electrically chiral quarks do not end up in fermionic bound states, but only bosonic ones, similar to six-quark states in QCD. If that is possible at all
we end up with a theory that does not have fermionic, charged baryons (``protons"), but does have charged bosonic nuclei. We would classify such a theory as ``acceptable".

However, the second argument for chiral symmetry breaking makes this scenario extremely implausible. 
This argument is the anomaly matching condition due to 't Hooft \cite{tHooft:1980xb}. He argued that one may weakly gauge the
entire flavor group. Because the vector representation of $SU(F)$ is anomalous this might seem inconsistent, but color singlets coupled to flavor can always be added to cancel that anomaly. 
These ``spectators" are not influenced by the $SU(M)$ dynamics. Whatever the color group does, it does not have the option to generate a massive spectrum of hadrons  and leave the entire
flavor group intact, because then the spectators would be left with uncancelled anomalies.  The possibilities are that either part of the flavor group is spontaneously broken, or that the anomalies
are matched by chiral massless baryons (in other scenarios the color group itself might also be broken). The hope was that there existed solutions of the latter kind, so that the chiral massless baryons could play the r\^ole  of composite quarks or leptons.

In the case of $SU(M)$ with $F$ quark-anti-quark pairs this anomaly matching requirement gives a strong argument in favor of chiral symmetry breaking, especially if one assumes 
$F$-independence. There is simply no plausible set of bound states to match the spectator anomalies. However, there still is one small caveat in applying this reasoning  here: part of the flavor group is already gauged (as it is in the Standard Model). It is gauged at least by electromagnetism, and in other examples also by an additional non-abelian group. Furthermore, it would be
wrong to assume that these gaugings are weak. Electromagnetism may be weak in our Universe, but we should not assume that about other universes. 

However, even in the presence of strong gaugings
there is  still an anomaly matching argument to be made. First of all, it may well happen that electromagnetism is not only chirally embedded in the flavor
group, but also anomalous, when restricted to the strong sector. In that case we already know that electromagnetism must be broken, or its anomalies must be matched by chiral baryons. That would be
fatal in itself. Massless charged baryons are equally bad as massless charged leptons. If electromagnetism is not anomalous in the strong sector, one may consider
the subgroup of the full flavor group that commutes with it, and with any other flavor gauge group. This included in any case all the $U(1)$'s corresponding to representation
multiplicities,  combined in such a way that they do not have anomalies with respect to $SU(M)$ color. Most of these have anomalies with respect to $U(1)_{\rm em}$. These anomalies
cannot be matched by chiral baryons, because then those baryons would have to be charged, and that is not acceptable. This leaves the possibility that all these chiral $U(1)$'s are
spontaneously broken, but that the chiral (but not anomalous) electromagnetic $U(1)_{\rm em}$ somehow survives, and is realized in the spectrum in the form of massive bosonic hadrons.

To see if there are any possibilities for this to happens requires a detailed case-by-case inspection, which we will not attempt. In the rest of the paper we will simply assume that
$SU(M)$ gauge theories with $F$ quarks and anti-quarks break their chiral symmetries to a vector-like $SU(F)$. There may remain isolated cases, presumably with very strong electromagnetic
coupling, where a different solution is possible. 

If there are other chiral color multiplets than just vectors we enter into much less understood territory. There is no empirical argument, but the anomaly matching argument still holds.
Also in this case it is likely that chiral $U(1)$ flavor symmetries must either be broken, or the spectrum must contain massless chiral fermionic bound states, both of which are
unacceptable. We will see that in the class of theories we consider, we can rely on the strongest form of the argument if the strong interaction group is $SU(3)$, and if $U(1)_{\rm em}$ gets contributions from both brane stacks. 
Then the
lepton sector by itself forces chiral sextets to be absent, so that we are left only with triplets and anti-triplets. For other color groups this does not hold. Then the best we can do
is to make the stronger assumption that a chiral $U(1)_{\rm em}$ is broken, even in the presence of chiral rank-2 tensors.  Of course, the chiral tensors do not ameliorate the situation in any
way, they just make the argument less rigorous.

\subsection{Summary}\label{Summary}

Our assumptions can be summarized as follows
\begin{enumerate}
\item{The high energy gauge group is realized in some fundamental theory that restricts the choice of massless particle representations and enforces anomaly cancellation. }
\item{The high energy gauge group is broken to the low energy gauge group by a single Higgs boson, which is in one of the allowed representations in point 1.}
\item{With the exception of that Higgs boson, all matter is chiral with respect to the high energy gauge group.}
\item{After the Higgs mechanism, the low energy gauge group contains at least one unbroken $U(1)$.}
\item{After the Higgs mechanism, all charged free leptons are massive, and the non-abelian interactions do not break the $U(1)$.}
\end{enumerate}
The last criterion can be broken up into a variety of stronger and weaker possibilities, as follows
\begin{itemize}
\item{5a. After the Higgs mechanism, all charged free leptons are massive, and the non-abelian interactions leave an unbroken $U(1)$.}
\item{5b. After the Higgs mechanism, all remaining gauge-coupled matter is non-chiral.}
\item{5c. After the Higgs mechanism, all remaining gauge-coupled matter is massive.}
\end{itemize}
The first possibility, 5a,  is what is really necessary, but it is hard to check in general because it requires a detailed understanding of chiral gauge theories. The second one
is much easier to check, but slightly too strong. The third one is realized in the Standard Model, but not really necessary for quarks. It one makes this assumption,
one can already dismiss many cases where particles do not couple to the would-be Higgs boson. Showing that the Standard Model is a more or less unique solution to
any of these conditions is interesting in its own right, even if a convincing anthropic argument is not immediately obvious. 
But in this paper we will just impose condition 5a, which does have a convincing anthropic motivation. 

Ideally, one could proceed as follows. Take any of the allowed  representations as the Higgs representation. 
 Then consider all possible symmetry breakings this Higgs allows. This leads to a low-energy gauge group consisting of one or more non-abelian factors and some $U(1)$'s.
 Then analyze the dynamical symmetry breaking induced by the entire non-abelian gauge group. One may also consider the Higgsless option, in which
 case only the last step needs to be considered. The resulting spectrum can be rejected if it does not contain a massless $U(1)$ or if it does contain massless charged leptons.
 
The problem is that we can only analyze dynamical symmetry breaking in a very limited number of cases.
Therefore in practice we will eventually have to make a  concession
 and add
one more assumption, namely that there is one strong interaction gauge group that is unaffected by the Higgs mechanism and dominates
the infrared dynamics. In the brane set-up we use, this r\^ole will be played by the $SU(M)$ subgroup of the first stack. We will assume
that any non-abelian gauge interaction from the second brane stack can be ignored. It may be either broken by the strong $SU(M)$ interaction
or just provide a next stage of non-abelian bound state dynamics. We will not examine if it breaks $U(1)_{\rm em}$ or replaces it by
another linear  combination.

\section{Higgs Multiplets}\label{HiggsMultiplets}

In this section we discuss a fact that is not as extensively discussed, namely that 
there is something really special about the way the Higgs mechanism works in the Standard Model. It has the unique property that each quark and lepton
gets its mass from just a single coupling to the Higgs and its conjugate. Although this feature will not play a central r\^ole in the discussion, it is noteworthy.

Consider the weak gauge group $SU(2)\times U(1)$ and the Standard Model Higgs boson $H$ in the representations $(2,-\frac12)$. In the Standard Model
this couples the two charged fermions\footnote{Generically, there  are two charged fermions; the lepton doublet in the Standard Model is an exception to this rule.} in a left-handed doublet to two left-handed singlets of opposite electric charge. This requires one Yukawa coupling to $H$ and another  to $H^*$. 
If we try to generalize this to other weak $SU(N)$ gauge groups and/or other matter representations we find that the Standard Model configuration is unique. 

More precisely, consider a weak group $SU(N)\times U(1)$ and a charged Higgs boson $H$ in the $SU(N)$ vector representation $(R,q)=(N,h)$, where the second entry is the charge. This Higgs boson breaks the symmetry to $SU(N\!-\!1)\times U(1)$, in such a way that a vector representation with arbitrary charge $q$ decomposes
as $(N,q) \rightarrow (N\!-\!1,q+\frac{h}{N\!-\!1})+(1,q-h)$. Consider now a Weyl spinor $\Psi$. It can have Yukawa couplings to spinors in the tensor product representations 
$\Psi^* \otimes H$ and $\Psi^*\otimes H^*$ (we denote fields and representations with the same symbol). These tensor products contain several components. What is special about
the Standard Model is that the dimensions of the smallest components in the tensor products add up to the dimension of $\Psi$. This is a necessary condition for pairing all states
in the set $(\Psi, \Psi^* \otimes H, \Psi^*\otimes H^*)$ into massive Dirac fermions. Consider for example an $SU(2)$ representation with spin $j$ and a 
Standard Model Higgs.
 The Yukawa couplings can pair it
with fermions in the representations $j\pm \frac12$, but even if we take the smallest ones, there is no way to match the dimension of $\Psi$
for arbitrary values of $j$. The only solution to $2j+1=2 \times (2j)$ is $j=\frac12$.
For $N>2$ not even the vector representation for the fermions satisfies the analogous condition. 

This does not mean that one cannot give masses to higher representations. It turns out that one may consider larger sets of representations with the property that all fermions can acquire a mass
from the Higgs coupling. We will call such a set a ``Higgs multiplet". In $SU(2)$,  for arbitrary $j$,  it consists of the representations
\begin{equation}
\label{HiggsMult}
{\cal H}(j,q) \equiv (j,q)_{+} + (j-\tfrac12,q-\tfrac12)_{-} + (j-\tfrac12,q+\tfrac12)_{-} + (j-1,q)_{+}
\end{equation} 
where the Higgs boson representation is $(j,-\frac12)$, using the Standard Model normalization. The subscript denotes the chirality. If it is $-$, one should either
use a right-handed fermion in the given representation, or a left-handed one 
with the complex conjugate of the representation.  We choose the second option, since we are using left-handed fields throughout. 
There are Yukawa couplings combining the first entry with the second, and the first     with the third, and yet another Yukawa coupling combining the third with
the fourth. 

A Higgs multiplet can be combined with a representation
of another gauge group, for example color. Then one must assign all members of the Higgs  multiplet to the same representation of that group, and
conjugate the entire representations according to the chirality subscript.

The simplest Higgs multiplet in $SU(N)$ is
\begin{eqnarray}
\label{VectorHiggsMult}
([0],q-h)_{-}+([1],q)_{+}+([2],q+h)_{-},\nonumber \\
\ldots, ([r],q+(r-1)h)_{\epsilon(r)} \ldots ([N],q+(N\!-\!1)h)_{\epsilon(N)}
\end{eqnarray}
where $[r]$ denotes an anti-symmetric tensor of rank $r$ and the Higgs is $([1],h)\equiv (V,h)$. The subscript $\epsilon(r)$ denotes the chirality, which must be alternating, $\epsilon(r)=-(-1)^r$.
The Higgs multiplet for a rank-2 tensor in $SU(N)$ is depicted in fig. \ref{HiggsMult} in terms of Young tableaux. The vector Higgs multiplet (\ref{VectorHiggsMult}) can be depicted in a similar way. It corresponds
to the first column of fig. \ref{HiggsMult} but with an additional singlet on top with an arrow pointing downwards to the box at the top of the column. We denote these two multiplets
as ${\cal H}(V,q)$ and ${\cal H}(T,q)$.

\begin{figure}[h!]
\begin{center}
\includegraphics[width=5.5cm]{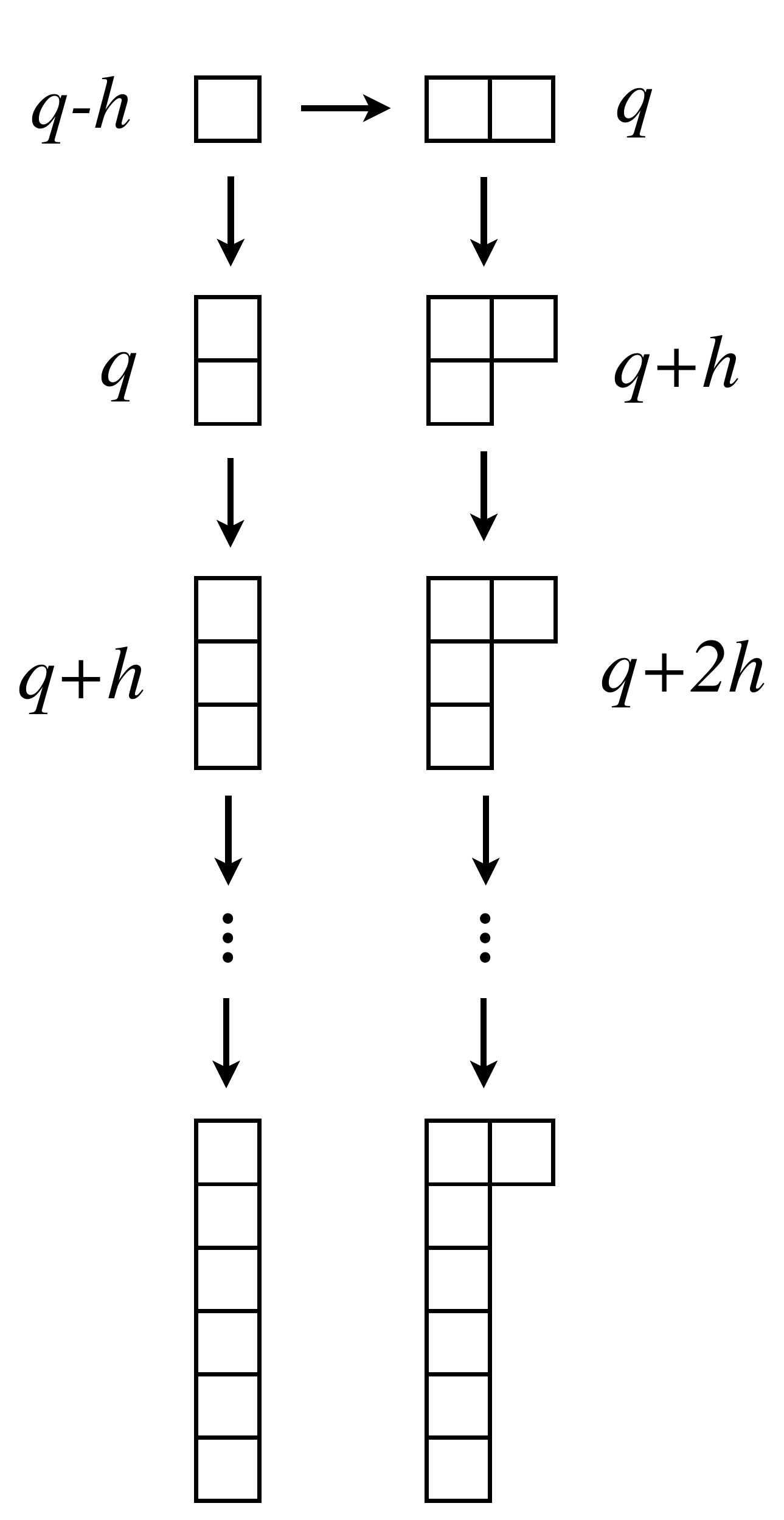}
\caption{\label{HiggsMult}\small Higgs Multiplet for rank 2 tensors for $SU(N) \times U(1)$ with a Higgs (V,h).}
\end{center}
\end{figure}

Each arrow corresponds to a Yukawa coupling of the form 
\begin{equation}
H(V,h) \Psi_L(R_1,q) C \Phi_L(R_2^*,-q-h)
\end{equation}
 where  $(R_1,q)$ denotes the $SU(N)\times U(1)$ representation where
the arrow starts and $(R_2,q+h)$ the representation where it ends, and $C$ is the charge conjugation matrix in spinor space needed to couple two left-handed fermions to a Lorentz singlet.
The field $\Phi$ must be in the representation conjugate to $(R_2,q+h)$, and must appear in the spectrum with the same multiplicity as $\Psi$. Indeed, all fields in the multiplet must appear
with the same multiplicity in the spectrum, but some field multiplicities may be distributed over more than one multiplet. For example if a spectrum can be written as  
$P\times {\cal H}(V,q\!-\!h)+Q\times {\cal H}(T,q)$, then there are $P\!+\!Q$ anti-symmetric tensors  and $Q$ symmetric tensors with charge $q$.

Each node in the diagram ({\it i.e.} each  Young tableau)
must have the property that all Higgses coupling to it must give mass to all components in the field. We have only checked that at least all
$SU(N-1)\times U(1)$ representations can be  paired, but not that in all cases a non-zero mass is actually generated for all components.  

Since we will encounter symmetric tensors of at most rank 2
we do not need anything more. 
However, we did work out the generalization of ${\cal H}(T,q)$ to symmetric tensors of rank $k$. It is the same as fig. \ref{HiggsMult}, but with a row of $k-2$ boxes 
attached to the top row of each Young tableau. It is an interesting mathematical
problem to write down such multiplets for arbitrary gauge groups and Higgs field choices, but we will not explore this issue any further here.  To generalize this result, 
the definition of Higgs multiplets needs to be made
more precise. Furthermore one has to establish a notion of reducibility and uniqueness and demonstrate the existence of a finite Higgs multiplet for a given field. Note that the tensor product of a Higgs field
with a fermion can yield several representations, so it is not excluded that there will be cases where nodes have more than one arrow. It is also possible that in some cases the procedure does
not converge within a finite set of fields.
Note that in general a Higgs multiplet depends not only
on the choice of the Higgs representation, but also on the subgroup the Higgs selects, which is not in all cases unique. The Standard Model is special in the sense that only one coupling to $H$ and
one to $H^*$ is needed to generate masses for all fermions in the multiplet.

For special values of $q$ the representations in a multiplet can be self-conjugate or each others' conjugates.
An example is the Standard Model
Higgs multiplet  ${\cal H}(\frac12,-\frac12)$, which does not need the right-handed neutrino at the end of the chain. Another example
occurs for $N=3$. A multiplet ${\cal H}(V,q)$ consists of the representations
\begin{equation}
\label{SUTHREEmultiplet}
{\cal H}(V,q)=(1,-q+h)+(V,q)+(V,-q-h)+(1,q+2h)
\end{equation}
If $q=-\frac12 h$ this becomes $2\times [(1,\frac32 h)+ (V,-\frac12 h)]$, so that one can get a massive multiplet with only two components
instead of four. Note that for $N=3$ a Higgs multiplet contains two vectors, and not a vector and an anti-vector. For $N=4$ we get 
\begin{equation}
\label{SUFOURmultiplet}
{\cal H}(V,q)=(1,-q+h)+(V,q)+(A,-q-h)+(\overline{V},q+2h)+(1,-q-3h)\ ,
\end{equation}
where $A$ is a rank-2 anti-symmetric tensor. In this case the beginning and the end of the chain are non-chiral for $q=h$ and $q=-3h$ 
respectively, and the entire multiplet is non-chiral for $q=-h$.

One can build alternatives to the Standard Model where all fermions get a mass from the Higgs by combining several Higgs multiplets into anomaly-free combinations. One may add one or several
``color" groups, and choose any representation of the group and assign it to each Higgs multiplet. The color and color-charge mixed anomalies cancel automatically within each Higgs multiplet, and so does 
the gravitational $U(1)$ trace anomaly. This is true for all Higgs multiplets that can be constructed, because it follows from the fact that all fermions can be chirally paired when the symmetry is broken.

In general, the $SU(N)$ and $U(1)$ cubic and mixed anomalies must be canceled among different Higgs multiplets (note that the cubic $SU(N)$ anomalies cancel within the Higgs multiplet given above 
for $N$ even, but not for $N$ odd, as one can see in the examples above). 
Consider now $N=2$. One can easily work out the two non-trivial chiral anomalies. The cubic anomaly of the multiplet is $-3jq$.  The $SU(2)^2 \times U(1)$ mixed anomaly is $2jq$.
Both anomalies cancel in any combination satisfying $\sum_i N_i j_i q_i=0$, where $i$ labels different Higgs multiplets with highest weak isospin $j_i$, charge $q_i$ and multiplicity $N_i$. 
The Standard model solves these conditions with $N_1=3, j_1=\frac12, q_1=\frac16$ and $N_2=1, j_2=\frac12, q_2=-\frac12$, per family. In terms of Higgs multiplets a Standard Model family
takes the form 
\begin{equation}
[3,{\cal H}(\tfrac12,\tfrac16)]+[1,{\cal H}(\tfrac12,-\tfrac12)]
\end{equation}

One often hears the claim that anomaly cancellation
fixes the electron-proton charge ratio to exactly $-1$. This is true in the sense that changing one of the charges $q_1$ and $q_2$ by any amount destroys anomaly cancellation, but from a more
general perspective one can easily write down many solutions to $\sum_i N_i j_i q_i=0$, including solutions with irrational charges. This is true even though the spectrum is chiral and couples to the
Higgs. Therefore this is {\it not} a solution to the charge quantization problem.

The minimal anomaly-free solution requires two Higgs multiplets, but if one chooses $q_1=-q_2$ the entire configuration is non-chiral. The simplest solutions that are non-chiral are given by the set
of generalizations of the Standard Model with $SU(M)$ color, 
\begin{equation}
[M,{\cal H}(\tfrac12,\tfrac1{2M})]+[1,{\cal H}(\tfrac12,-\tfrac12)]
\end{equation}
One may also obtain a non-trivial solution by using higher isospin $SU(2)$ representations, for
example ${\cal H}(1,1)+{\cal H}(\tfrac12,-2)$.

\section{Towards a derivation of the Standard Model}\label{Derivation}

Our chirality assumptions immediately rule out a pure, single photon electromagnetic theory.
The fermion spectrum can be chosen chiral, but then to get a non-chiral
low energy spectrum the Higgs has to break the $U(1)$. 

Hence we need to extend the $U(1)$ with additional gauge symmetries. In quantum field theory the number of possibilities for a chiral gauge theory
that is broken by a single Higgs to a non-chiral spectrum is gigantic, and at this point we are going to need a well-motivated top-down assumption, 
namely a particular string realization of our theory.

\subsection{Single stack models}\label{SingleStack}

We will consider brane realizations, starting
with the minimal number of branes. The allowed particles in the massless spectrum in such models are  vectors, adjoints, symmetric and 
anti-symmetric tensors, all belonging to a single brane stack, and bi-fundamentals stretching between stacks.  We allow for open strings with just one endpoint
on the stack, and another endpoint on a neutral object. This might for example be a $O(1)$ brane or a $U(1)$ brane with an anomalous gauge
symmetry, so that the gauge boson gets a mass from a Green-Schwarz mechanism. 
These give rise to vectors on the stack to which the other end of the string is attached.

The first option to consider is that the electromagnetic $U(1)$ is embedded in a single brane stack. This must be a unitary stack, since otherwise all representations are non-chiral. The gauge group
is $U(N)$.
The spectrum of the single stack consists of $K$ vectors of charge $q$,  $S$ symmetric tensors of charge $2q$ and  $A$ anti-symmetric tensors  of charge  $2q$, where
the charge refers to the overall phase $U(1)$ of the stack. 
Here  $K,S$ and $A$ can be both positive and negative (a sign change implies a chirality change), and $q$ can be zero -- if the $U(1)$ is broken by the Green-Schwarz mechanism -- or non-zero.
The $U(1)$ has to satisfy the following cubic and mixed (gauge and gravity) anomaly cancellation conditions
\begin{eqnarray*}
K N q^3 + \tfrac12 N(N+1) S  (2 q)^3+ \tfrac12 N(N-1) A  (2 q)^3 &=&0\\
K N q + \tfrac12 N(N+1) S  (2 q)+ \tfrac12 N(N-1) A  (2 q)&=&0   \\
K q+(N+2) S (2q) + (N-2) A (2q) &=& 0
\end{eqnarray*}
These can only be solved if either $K=S=A=0$ or $q=0$. In the latter case $K, S$ and $A$ are only constrained by  cubic $SU(N)$ anomaly cancellation, and a 
chiral spectrum can be obtained. But then the electromagnetic $U(1)$ must  emerge from a Higgs  breaking $SU(N)$.  The choice of Higgs bosons is:
a vector, a symmetric tensor, an anti-symmetric tensor or an adjoint. The resulting symmetry breaking patters have been worked out in \cite{Li:1973mq} (with a correction
in \cite{Elias:1975yd}).
A vector breaks $SU(N)$ to $SU(N\!-\!1)$, a symmetric tensor breaks it to $SO(N)$ or to $SU(N\!-\!1)$ (depending on the Higgs potential) and
an anti-symmetric tensor breaks $SU(N)$ to $Sp(N)$ (if $N$ is  even) or $Sp(N\!-\!1)$ (if $N$ is odd), or to $SU(N\!-\!2)\times SU(2)$. The only way these symmetry breakings could yield a $U(1)$  is if
$SU(2)$ is broken by means of a symmetric tensor to $SO(2)$. But $SU(2)$ has no complex representations, and hence is not a suitable high-energy theory by itself; it violates assumption 3. An adjoint representation breaks $SU(N)$ to $SU(p)\times SU(q)\times U(1)$, $p+q=N$. This looks promising, because at least
it produces a $U(1)$. But it is easy to see that this can never break a chiral representation to a non-chiral one. We will discuss this
in more detail for two-stack models in section \ref{HiggsChoice}.


\subsection{Two Stack Models}

The next possibility is to obtain the $U(1)$ from two brane stacks.  In this paper we will only consider the possibility that both are unitary, and consider a general 
$U(M)\times U(N)$ two-stack model.
The gauge group is $SU(M)\times SU(N) \times U(1)^2$, but anomalies (canceled by a Green-Schwarz
mechanism) will leave at most one linear combination of the two $U(1)$'s unbroken. We will write it as $Y=q_aQ_a+q_b Q_b$ where $Q_a$ and $Q_b$ are
the brane charges of the two stacks. The possibilities for chiral matter representations are then (note that adjoints are not chiral, so we do
not have to consider them)
\begin{eqnarray}
\label{BraneReps}
Q&(M,N,q_a+q_b)\nonumber\\
U&(A,1,2 q_a)\nonumber\\
D&(\overline{M},1,-q_a)\nonumber\\
S&(S,1,2 q_a)\nonumber\\
X&(M,\overline{N},q_a-q_b)\\
L&(1,\overline{N},-q_b)\nonumber\\
T&(1,S,2 q_b)\nonumber\\
E&(1,A,2 q_b)\nonumber
\end{eqnarray}
where $A,S$ denote (anti)symmetric tensors. We have given these multiplets suggestive names referring to the Standard Model, but of course those names can correspond to genuine quarks and leptons
only for $M=3$ and $N=2$. 
We  will use variables $Q,U,D,\ldots$, which can be any integer, to denote the multiplicity
of  these representations.  If a multiplicity  is negative this implies a positive multiplicity for the conjugate representation. The representations themselves will be denoted as ${\bf Q},{\bf U},{\bf D},\ldots$. 
We have chosen to use the anti-vectors for {\bf L} and {\bf D}, because then the Standard Model multiplicities will be positive integers. Note however that for notational
convenience we have not added superscripts to denote anti-particles. So ${\bf U}$ and {\bf D} correspond to anti-quarks in the Standard Model,
and {\bf L} corresponds to anti-leptons.

\subsubsection{Anomaly cancellation conditions}

The integer multiplicities are subject to anomaly cancellation. We will denote anomalies by a three-letter code, where `S',`W' and
`Y' refer to $SU(M)$, $SU(N)$ and $U(1)$, and `G' to gravity.
Hence we have anomalies of type SSS, SSY, WWW, WWY, YYY and GGY. Note that the WWW anomaly is trivial in field theory for $N=2$,  but in a brane model the
requirement of tadpole cancellation still imposes it as if it were a non-abelian anomaly. Hence the anomaly  contributions of vectors, symmetric and anti-symmetric tensors are 1, $N+4$ and $N-4$ respectively, even
for $N=2$ (the case $N=1$ is discussed below).
We will see however that  there is a linear dependence among the six anomalies, so that the WWW anomaly is not really needed. Since we want to assume as little
as possible about the string theory origin of these gauge groups, it is useful to know that the anomalies we use are really just the field-theoretic ones. Furthermore, we can use the linear
dependence to trade the awkward YYY anomaly for the much more manageable WWW anomaly.

The condition
of anomaly cancellation constrains the  parameters $q_a$ and $q_b$ as well as the particle multiplicities. Note that in brane models, $U(1)$'s do not have to be anomaly free, because their
anomalies are canceled by the Green-Schwarz mechanism. But in that case the corresponding gauge boson acquires a mass, and cannot be the one of the Standard Model.
In brane models it may also happen that a non-anomalous $U(1)$ acquires a mass from mixing with axions, but this is irrelevant for our purposes. There exist models where this
is not the case, and those are the only ones of interest. 


The anomaly cancellation conditions can be greatly simplified and brought to the following form
\begin{eqnarray}
\label{Aneq}
(S+U) \tilde q_a &=& C_1\nonumber \\
 (T+E) \tilde q_b   &=& -C_2 \nonumber\\
(D + 8 U) \tilde q_a  &=&   (4+M) C_1 + N C_2 \\
L \tilde q_b + D \tilde q_a &=& 0 \nonumber\\
 2 E \tilde q_b + 2 U \tilde q_a  &=& C_1-C_2\nonumber
 \end{eqnarray} 
 Here $\tilde q_a \equiv M q_a$,  $\tilde q_b \equiv N q_b$, $C_1=-(Q-X) \tilde q_b$ and $C_2=(Q+X)\tilde q_a$. The Standard model parameter values are $\tilde q_a=-1$, $\tilde q_b=1$,
 $C_1=C_2=-3$, $Q=U=D=L=E=3$ and $S=T=X=0$, and of course satisfy these equations for $M=3$, $N=2$. For any $M$ and $N$ there are just five independent equations, demonstrating
 that the WWW equation is redundant even if $N\not=2$.  
 
 In the derivation of these equations we used $N\not=1$, $M\not=1$,
  $q_a\not=0$ and $q_b\not=0$. If $N$ or $M$ are equal to one, the SSS and WWW anomaly conditions continue to hold in a brane model, because they follow from the requirement of tadpole cancellation.
  If $N=1$ this leads to the strange results that the open string sector $E$ contributes to anomaly cancellation, even though it contains no massless states!   
  However, the reason (\ref{Aneq}) is not necessarily valid is that the SSY and/or WWY anomaly cancellation conditions have no meaning anymore if $M$ and/or $N$ are equal to 1.

  If we  choose just one of the two brane stack multiplicities equal to one, we lose one equation, but we still have five left. Since the original set of six equations has a redundancy, one may expect
 to obtain exactly the same equations, and by inspection this is indeed correct. Note that for $N=1$ or $M=1$ the anomaly cancellation conditions are not just the field theoretic ones, but
  that there is one stringy SSS or WWW condition.
  
  The special cases $q_a = 0$ or $q_b=0$ can be included by adding the $SU(M)$ and $SU(N)$ anomaly cancellation conditions $(M+4)S+(M-4)U-D+2NQ=0$ and $(N+4)T+(N-4)E-L+2MQ=0$ to the set.
  These conditions are redundant if both charges are non-vanishing, and can be obtained from (\ref{Aneq}) after dividing an equation by $q_a$ or $q_b$ and then taking the limit $q_a=0$ or $q_b=0$. 
  No matter which case one considers, all seven
  equations reduce to five independent ones. For $q_a=0$ the equations reduce to the $SU(M)$ anomaly cancellation condition, $X=Q$ and $L=T=E=0$. For $q_b=0$ the solution is $X=-Q$ and $U=D=S=0$, with $L, T, E, Q$
  and $X$ subject to the $SU(N)$ anomaly cancellation condition.
  If both $q_a$ and $q_b$ vanish, we only get two equations, the $SU(N)$ and $SU(M)$ anomaly cancellation conditions.

  Note that the  equations (\ref{Aneq}) are invariant under the interchange of $X$ and $Q$,
and sign changes of $L, T$, and $E$ and $\tilde q_b$. So we may assume, without loss of generality, that $Q \geq X$. Furthermore 
by fixing the overall chirality we may assume that $Q \geq 0$.

 \subsubsection{$U(1)\times U(1)$}\label{UoneUone}

 The anomaly cancellation conditions (\ref{Aneq}) are not valid if $M=N=1$.
 In this case we only have SSS, WWW, GGY and YYY anomalies, but no mixed gauge anomalies. This assumes that there is
 just a single $U(1)$ factor, $Y$.
 But $U(1)\times U(1)$ theories are only of interest if {\it both} $U(1)$'s are anomaly free, since we have already ruled out a 
 pure $U(1)$ theory on general grounds. 
 It is easy to see that having two $U(1)$'s  is not possible.  This can be proved without using the stringy SSS and WWW anomalies. The ${\rm Y}_1{\rm Y}_1{\rm Y}_2$ and ${\rm Y}_1{\rm Y}_2{\rm Y}_2$ mixed anomalies only get contributions from
 {\bf Q} and {\bf X}, and cancel if $Q=\pm X$, hence $Q=X=0$. The $GG{\rm Y}_1$ anomalies in the first factor cancel if $-D+2S=0$, and the cubic  ${\rm Y}_1{\rm Y}_1{\rm Y}_1$ anomalies cancel if $-D+8S=0$. Hence 
 $D=S=0$. Note that $U$ and $E$ have no massless physical states. Similarly, in the second factor $L=T=0$. Hence there are no massless states. If only one $U(1)$ is anomaly free and chiral, the Higgs mechanism
must break it completely. 
 
Hence $U(1)\times U(1)$ yields no solutions. This implies that there {\it must} be at least one non-abelian factor in the high energy gauge group. At this point we will make an additional assumption, as explained in section \ref{Summary}, namely that this non-abelian factor plays the r\^ole
of the strong interactions. In particular, will assume that  it is not broken
by the Higgs mechanism. 

 \subsubsection{Higgs choice}\label{HiggsChoice}

In principle all of the representations in (\ref{BraneReps}) are available as chiral matter representations, but also as Higgs representations. 
The choices that leave $SU(M)$ unbroken are {\bf L}, {\bf T} and {\bf E}.
But 
we should  also allow the adjoint representation, which is not taken into account as a fermionic  matter representation because it is real.
However, we will now show that the adjoint is not going to do the job of turning a
chiral spectrum into a non-chiral one. 

Consider the adjoint Higgs of $SU(N)$, first with $N > 2$.  Then $SU(N)$ breaks to
$SU(p)\times SU(q)\times U(1)$, where we choose $p \geq q$. 
Hence $p$ must be at least 2. A vector representation breaks to $(V,1,q)+(1,V,-p)$.
If the spectrum contains
a chiral tensor  {\bf T}, it contains  a chiral tensor of $SU(p)$ with charge $2q$. This has no match in the spectrum, so $T$ must vanish.
Similarly,  {\bf E} contains an anti-symmetric tensor with charge $2q$, which even for $p=2$ (when the tensor is a singlet) cannot be matched.
Hence $E$ must vanish as well. A field {\bf L} contains $SU(p)\times SU(q)\times U(1)$ representations that can only be matched by $Q$
or $X$, and only for $M=1$. If $M=1$ one may be able to pair off components of {\bf Q}, {\bf X} and {\bf L}, but only if $q_a=0$. But then
${\bf Q}^*$, {\bf X} and {\bf L} are the same representation, and if they can pair off after Higgs symmetry breaking, they can 
also do so without Higgs breaking. For all other values of $M$ we must have $T=E=L=0$. Furthermore {\bf Q} and {\bf X} must pair off,
which is only possible if $M=2$, $Q=X$ and $q_a=0$. Otherwise {\bf Q} and {\bf X} must also vanish, and the entire
solution is trivial. If $M=2$, $Q=X$, $T=E=L=0$ and $q_a=0$, the entire spectrum is non-chiral before symmetry breaking. 

This leaves the case  $N=2$ to be discussed. 
An adjoint Higgs breaks $SU(2)$ to $SO(2)$. The reality properties of $SU(2)$ and $SO(2)$ are nearly identical, so if a mass term is
allowed after symmetry breaking, it will also be allowed without symmetry breaking, and the Higgs does not have any effect. The only
subtlety here is the pseudo-reality of $SU(2)$. For example, suppose $q_b=0$. Then the $Y$ charge does not inhibit a mass term for {\bf L}.
However, such a mass term  must be anti-symmetric in the flavor index of {\bf L}, 
and hence if $L$ is odd there is generically a single field {\bf L} that
remains massless. If $SU(2)$ is broken
to $SO(2)$,  there is an allowed  Higgs coupling $H{\bf L}{\bf L}$ that can give a small mass to the remaining massless field {\bf L}.
Hence if
$L$ is odd this would generically give rise to a single light charged lepton pair with
opposite charges. The other fields where we can encounter pseudo-reality of $SU(2)$ doublets are {\bf Q} and {\bf X}. This can happen
only for self-couplings of those fields, so we must have $q_a=\pm q_b$, and $M < 3$. If $M=2$ the self-coupling is anti-symmetric
in both color and flavor, so that there is no problem. This leaves $M=1$. All these cases [$q_b\not=0$ and $L$ odd; 
$q_a=q_b, M=1$ and $X$ odd; $q_a=-q_b, M=1$ and $Q$ odd] lead to the same answer. 
All other fields can acquire at large mass before symmetry breaking
they can do so after symmetry breaking. Only a single field {\bf L}, {\bf X} or {\bf Q} remains. It gets a small mass from the adjoint Higgs.
It seems that we have here a solution to our conditions where the only light field is a single charged Dirac fermion, 
like an electron-positron pair.
But if there is a single unpaired $SU(2)$ doublet in the unbroken theory,  it has a global
Witten anomaly \cite{Witten:1982fp}, and hence it is inconsistent.

 \subsubsection{$U(3)\times U(1)$}\label{UthreeUone}

  If $N=1$ the Higgs {\bf E} has no massless states, and the Higgses {\bf L} and {\bf T} are charged singlets,
 which will break the $Y$ charge if they get a vev, and if $q_b\not =0$. Since getting a massless $U(1)$ is our main objective, this is not acceptable.  
  
 If $q_b=0$ all Higgs candidates are singlets and cannot break any symmetries.  From the
 anomaly cancellation conditions we get $U=D=S=0$, $X=-Q$, so that the $SU(3)$ spectrum is $(3,q_a) + (\overline{3},-q_a)$. This is non-chiral, and hence will not be present among
 the massless states. All leptons have charge zero and hence there are neither light quarks nor light charged  leptons.

\subsubsection{$U(3)\times U(2)$}\label{SMcase} 
 
 So let us move to the next case, $N=2$. For any $N$, one could argue that  sextets can be ruled out because they cannot couple to any of the candidate Higgses, and
 hence remain chiral and massless. However, as discussed earlier, we consider the requirement of a non-chiral spectrum for the non-abelian groups only as a last resort, and therefore it is better to rule
out {\bf S} by a different argument. For $M=3$ and $N=2$ this can indeed be done. First we discuss the special cases  $q_b=0$ or $q_a=0$.
The case $q_a=q_b=0$ is discussed for general $M$ and $N$ in subsection \ref{qaisqbiszero}.

The case $q_b=0$ works in a similar way for $N=2$ and $N=1$.
Anomaly cancellation requires $U=D=S=0$, $X=-Q$, so that the $SU(3)$ spectrum is $(3,2,q_a) + (\overline{3},2,-q_a)$. This is non-chiral, and hence massive by assumption 3. All leptons have charge zero and hence they are non-chiral and massive as well. Note that for $q_b=0$ and $N=2$
the Higgs boson is equivalent to an adjoint. Then the discussion in section \ref{HiggsChoice} applies, and there is no solution.

If $q_a=0$ we have $L=T=E=0$, $Q=X$, $7S-U-D+4Q=0$. The bi-fundamentals contribute $Q\left[(3,2,q_b)+(3,2,-q_b)\right]$. Their $SU(3)$ anomaly is canceled by uncharged symmetric tensors
and anti-triplets. A Higgs in the {\bf E} representation breaks the $U(1)$, and a Higgs in the {\bf L} breaks $SU(2)\times U(1)$ in the standard way, so that the resulting spectrum is  
$Q\left[(3,2q_b)+(3,-2q_b)+2(3,0)\right]$. This $U(1)$ is chiral (the photon only couples to left-handed quarks), and will be broken by QCD in theories with $S=0$.  A Higgs {\bf T} in the triplet representation can either break $SU(2)\times U(1)_Y$ in the same way as {\bf L}, or it can break it to $SO(2)$, embedded only in $SU(2)$ 
(then $U(1)_Y$ is broken). The resulting spectrum is $Q\left[2(3,1)+2(3,-1)\right]$. Once again the photon only couples to
left-handed quarks, and the electromagnetic $U(1)$ will be broken by QCD if $S=0$. 

The presence of chiral sextets ($S\not=0$) complicates the discussion. They are not electrically charged, and
the photon still couples only to the left-handed triplets. Without the sextets, this would imply that QED is broken.
There is  no reason to believe that the presence of an uncharged chiral sextet will improve the situation, but on the other hand
we do not know the bound state spectrum in that case. 
The low energy spectrum is obviously chiral, so this does violate condition 5b and 5c, but we regard the violation  of condition 5a as
undecided.

 


So from now on we can assume $q_a\not=0$ and $q_b\not=0$.
The three possible Higgses ${\bf L}$, ${\bf T}$ and ${\bf E}$ are a doublet, a triplet and a singlet of $SU(2)$. 
\vskip .5truecm
\leftline{\bf Doublet Higgs}\vskip .2truecm\noindent
Consider first the option  that leads to our Universe, namely a $SU(2)$ doublet Higgs. With our conventions, the unbroken electromagnetic charge is $2q_b T_3 + Y$.
Fermions in the representation ${\bf L}$ yield chiral charged particles with charges $0,-2q_b$, whereas $T$ yields charges $0,2q_b,4q_b$, and
we can also have fermions in the representation ${\bf E}$ with charge $2q_b$. This spectrum can only be non-chiral if $L=E$ and $T=0$. If not, there will be massless, and even chiral, charged leptons with
catastrophic consequences. Plugging this into the anomaly equations (\ref{Aneq}) we find $E\tilde q_a=C_2$, $D\tilde q_b=C_2$, $U\tilde q_a=\tfrac12 (C_1+C_2)$ and $S\tilde q_a=\tfrac12 (C_2-C_1)$.
Now we substitute this into the third equation of (\ref{Aneq}), and obtain
\begin{equation}
(5-N) C_1 = M C_2
\end{equation}
For $N=2$ and $M=3$ this result implies that $C_1=C_2$, and hence $S=0$ (note that there is a second solution to the condition $C_2=C_1$, namely $M=4, N=1$, and we will see later what that implies). 
Hence to avoid chiral leptons for $M=3$ we must set $S=0$. Since the anti-symmetric tensor of $SU(3)$ is an anti-triplet we are now in the desirable situation of an $SU(3)$ gauge group with matter only in
the fundamental representation. 

We will present the rest of the argument without directly using the anomaly conditions (\ref{Aneq}), because this is more insightful, and the derivation of (\ref{Aneq}) is straightforward, but rather tedious.
The quark  multiplets split up in the following way
\begin{equation}
Q(3,q_a)+Q(3,q_a+2q_b)+X(3,q_a)+X(3,q_a-2q_b)-U(3,-2q_a)-D(3,q_a)\ ,
\end{equation}
where we have conjugated ${\bf U}$ and ${\bf D}$ in order to  have only triplets. We have to pair all these components. 
The first term can be paired with a component of ${\bf X}$ and 
with ${\bf D}$, without any constraints on charges. But the second component can only be paired with ${\bf U}$, since $q_b\not=0$.
 Hence if $Q\not=0$, we find the relation $q_a+2q_b=-2q_a$, {\it i.e.} $3q_a=-2q_b$, and $Q=U$. This charge relation
implies immediately that there is no partner for the second component of ${\bf X}$, so that $X$ must vanish. Then the first component of ${\bf Q}$ can
only pair with {\bf D}, and we get $D=Q$. If $Q=0$, we can apply the same reasoning to ${\bf X}$, with the result $3q_a=+2q_b$, and $X=U=D$. This is
just the solution with $X \leftrightarrow Q$ interchange that exists on general grounds.
If $Q$ and $X$
both vanish there is no solution, since $q_a\not=0$. 

All anomalies involving $SU(3)$ already cancel, and the quark contribution
to the $U(1)$ trace anomaly cancels by itself. The relation between the charges $q_a$ and $q_b$ is  the familiar one from $SU(5)$, and so we know that all
particles have their familiar charges. We choose the Standard Model  normalization conventions.
We get the following equations for $L,T$ and $E$
\begin{eqnarray*}
\hbox{SSY} & \frac12 Q -\frac12  L+ 4T = 0 \\
\hbox{GGY} & -L+3T+E=0 \\
\hbox{YYY} & -\frac34 Q -\frac14 L +3T+E=0
\end{eqnarray*}
which imply that $L=E=Q$ and $T=0$.
Note that the $SU(2)$ anomaly $3Q-L+6T-2E$ is not really needed, and follows from the others. We do not need to check that the Higgs does indeed give mass
to all quarks and leptons, because this is the Standard Model.

 \vskip .5truecm
\leftline{\bf Triplet Higgs}\vskip .2truecm\noindent
The triplet Higgs can break $SU(2)\times U(1)$ in two ways \cite{Li:1973mq}, depending on the signs of two terms in the Higgs potential. 
The Higgs vev can either take the form
\begin{equation}
\langle H \rangle =
\begin{pmatrix}
0 & 0 \\
0 & v 
\end{pmatrix}\ ,
\end{equation}
which breaks $SU(2)\times U(1)$ to $U(1)$. 
The other possibility is a breaking to $SO(2)$ due to a Higgs vev 
\begin{equation}
\langle H \rangle =
\begin{pmatrix}
v & 0 \\
0 & v 
\end{pmatrix}\ .
\end{equation}
This has an obvious generalization to symmetric 
rank-2 tensors of $SU(N)$,  with breaking patterns $SU(N)\rightarrow SU(N\!-\!1)\times U(1)$ and $SU(N)\rightarrow SO(N)$. If the Higgs field is charged
under an additional $U(1)$ factor, this $U(1)$ is broken in the second case. In the first case there is a generator of $SU(N)$ that commutes
with $SU(N\!-\!1)$ and can be combined with the $U(1)$ generator. This combination remains unbroken. 
The first pattern is the same  as for the
vector Higgs. However, the Yukawa couplings are different if $H={\bf T}$, and cannot produce the ${\bf L}{\bf E}$ coupling of the Standard Model.
The triplet Higgs can give a Majorana mass to the left-handed
neutrino, but without an additional doublet Higgs
the charged leptons remain massless and this case must be discarded. 

The other breaking pattern, $SU(2) \times U(1)\rightarrow SO(2)$, as well as the Higgs choice $H={\bf E}$ will be discussed for general $M$ and $N$ in section \ref{Exceptional}.

\subsubsection{$U(M)\times U(N)$, $M \geq 3$, $N > 2$}

It is now natural to explore other choices of $M$ and $N$. For $N > 2$ we encounter a new feature which 
complicates the discussion, namely that there will be a non-abelian factor in the flavor group after Higgs symmetry breaking.
 
 \vskip .3truecm
\leftline{\bf Higgs Symmetry Breaking Patterns}
\vskip .2truecm

For example, 
if the Higgs is {\bf L}, the symmetry is broken to $SU(M)\times SU(N\!-\!1) \times U(1)$. The $U(1)$ is
a linear combination of $Y$ and the
generator $T$ embedded in the vector representation of $SU(N)$: 
\begin{equation}
T={\rm diag} (\frac{q_b}{N-1}, \ldots, \frac{q_b}{N-1},-q_b)
\end{equation}
This implies the following decompositions:
\begin{eqnarray}
\label{VectorHiggs}
SU(M) \times SU(N) \times U(1) &\rightarrow& SU(M) \times SU(N-1)\times U(1) \ \ \ \ \hbox{(Higgs {\bf L} or {\bf T})}\nonumber\\
Q &\rightarrow& (V,V,q_a+\tfrac{Nq_b}{N-1})+(V,1,q_a)\nonumber\\
U &\rightarrow& (A,1,2q_a)\nonumber\\
D &\rightarrow& (\overline{V},1,-q_a)\nonumber\\
S &\rightarrow& (S,1,2q_a)\nonumber\\
X &\rightarrow& (V,\overline{V},q_a-\tfrac{Nq_b}{N-1})+(V,1,q_a)\\
L &\rightarrow& (1,\overline{V},-\tfrac{Nq_b}{N-1})+(1,1,0)\nonumber\\
T &\rightarrow& (1,S,\tfrac{2 Nq_b}{N-1})+(1,V,\tfrac{N q_b}{N-1})+(1,1,0)\nonumber\\
E &\rightarrow& (1,A,\tfrac{2 Nq_b}{N-1})+(1,V,\tfrac{N q_b}{N-1})\nonumber
\end{eqnarray}

Note that the Higgs {\bf T} can break $SU(N)\times U(1)$ in exactly the same way.  It may also break $SU(N)$ to $SO(N)$ without any
surviving $U(1)$. The Higgs {\bf E}  can also break $SU(N)\times U(1)$ in two different ways: to $SU(N-2)\times SU(2)\times U(1)$ for $N \geq 3$
and to $Sp(N)$ if $N$ is even or  $Sp(N-1)\times U(1)$ if $N$ is odd. In the latter case  $Sp(N-1)$ is a 
maximal subgroup of the $SU(N-1)$ flavor group obtained by breaking with the Higgs {\bf L}. 
The decompositions are as follows:
\begin{eqnarray}
\label{AsymTensorHiggs}
SU(M) \times SU(N)\times U(1)  &\rightarrow&SU(M) \times  SU(N-2)\times SU(2) \times U(1) \ \ \ \ \hbox{(Higgs {\bf E})}\nonumber\\
Q &\rightarrow& (V,V,  1    ,q_a+\tfrac{Nq_b}{N-2})+(V,1,2,q_a))\nonumber\\
U &\rightarrow& (A,1,  1    ,2q_a))\nonumber\\
D &\rightarrow& (\overline{V},1,    1  ,-q_a))\nonumber\\
S &\rightarrow& (S,1,  1    ,2q_a))\nonumber\\
X &\rightarrow& (V,\overline{V},  1    ,q_a-\tfrac{Nq_b}{N-2})+(V,1,  2   ,q_a))\\
L &\rightarrow& (1,\overline{V},  1    ,-\tfrac{Nq_b}{N-2})+(1,2,0))\nonumber\\
T &\rightarrow& (1,S,   1   ,\tfrac{2 Nq_b}{N-2})+(1,V,   2   ,\tfrac{N q_b}{N-2})+(1,1,   3   ,0))\nonumber\\
E &\rightarrow& (1,A,  1   ,\tfrac{2 Nq_b}{N-2})+(1,V,   2   ,\tfrac{N q_b}{N-2})+(1,1,1,0))\nonumber
\end{eqnarray}
The second symmetry breaking pattern
for $N$ odd is: 
\begin{eqnarray}
\label{Symplectic}
SU(M) \times SU(N) \times U(1) &\rightarrow&SU(M) \times  Sp(N-1)\times U(1) \ \ \ \ \hbox{(Higgs {\bf E})}\nonumber\\
Q &\rightarrow& (V,V,q_a)+(V,1,q_a+Nq_b)\nonumber\\
U &\rightarrow& (A,1,2q_a)\nonumber\\
D &\rightarrow& (\overline{V},1,-q_a)\nonumber\\
S &\rightarrow& (S,1,2q_a)\nonumber\\
X &\rightarrow& (V,{V},q_a)+(V,1,q_a-Nq_b)\\
L &\rightarrow& (1,{V},0)+(1,1,-Nq_b)\nonumber\\
T &\rightarrow& (1,S,0)+(1,V,N q_b)+(1,1,2N q_b)\nonumber\\
E &\rightarrow& (1,A,0)+(1,V,Nq_b)+(1,1,0)\nonumber
\end{eqnarray}
Note that the anti-symmetric tensor of $SU(N-1)$ decomposes into an anti-symmetric tensor of $Sp(N-1)$ and a singlet,
the symplectic trace. This singlet corresponds to the Higgs vev if $H={\bf E}$. Note also that for $N=3$ the breaking pattern (\ref{Symplectic}) is
identical to (\ref{AsymTensorHiggs}), because $SU(N-2)$ is trivial. It is also identical to (\ref{VectorHiggs}) apart from the 
charge assignment.
This is as expected, because in that case the anti-symmetric tensor is an anti-triplet of $SU(3)$, which has only
one breaking pattern. The different charge assignment is due to the fact that the Higgs {\bf E} has a different charge than {\bf L}.

There is no need to list the other two breaking patterns $SU(N) \rightarrow SO(N)$ and $SU(N) \rightarrow Sp(N)$, $N$ even, since they
are essentially the same as (\ref{BraneReps}). The only differences are that $q_b=0$, $\overline{V}=V$, and that the
symmetric and anti-symmetric tensors contain an additional singlet for $SO(N)$ and $Sp(N)$, respectively. This singlet
corresponds to the Higgs vev if $H={\bf T}$ or $H={\bf E}$.

\vskip .3truecm
\leftline{\bf Chiral Symmetry Breaking}
\vskip .2truecm
In some of these cases, $U(1)_Y$ is broken, whereas in others it is chiral with respect to $SU(M)$, {\it i.e.} $SU(M)$ vectors
and anti-vectors\footnote{Note that massless rank-2 tensors will in any case be chiral.} do not have opposite $U(1)_Y$ charges, and
hence cannot acquire a mass. Then one could simply
discard the theory, but that would be premature. One should still ask the question if the strong interactions could break the flavor group to
a suitable electromagnetic group $U(1)_{\rm em}$, as in the Higgsless Standard Model.
We will see that this is indeed possible in some cases, but then massless charged leptons remain in the spectrum, exactly as
in the Higgsless Standard Model.

A possible way out is to assume that a non-abelian subgroup of the flavor group that commutes with $U(1)_{\rm em}$ remains unbroken by
the strong interactions. 
This flavor group may couple to the leptons, and give them a dynamical mass, the way QCD does with quarks. However, if the strong interactions 
behave as expected, that cannot happen. Consider one of the non-abelian factors $G$: $SU(N-1)$, $SO(N)$, $SU(N-2)$, $Sp(N)$, $Sp(N-2)$ or
$SU(2)$ that occur in the breaking patterns described above. We denote the vector representation of these groups as $V$, and its dimension as $d$. 
The flavor group couples to the quarks as
\begin{equation}
Q (M,V)+X(M,\overline{V})
\end{equation}
In addition there will be representations {\bf U}, {\bf D} and {\bf S} that do not couple to the flavor group. Assume that
$G$ breaks to some subgroup $H$, such that the vector representation contains a 
(not necessarily reducible)
representation
$r_H$ of $H$. In the absence of rank-2 tensors, experience from QCD suggests that this group must be such that the theory will be
non-chiral when we replace $G$ by $H$. Since the rank-2 tensors do not couple to the non-abelian part of the flavor group, it is
plausible that this will also be true in the presence of rank-2 tensors.

If we consider $H$ instead of $G$, vectors from {\bf Q} are coupled to the $H$-representation $Q \times r_H$. For this to become
non-chiral, we need  anti-triplets from {\bf X} in the representation  $-X \times r_H$. The result is non-chiral only if $X=-Q$ and
$r_H$ is a real or pseudo-real representation. Then {\bf Q} and {\bf X} would be exactly paired, but then the presence of either {\bf U}, {\bf D} or {\bf S} would
eliminate any possibility of non-chiral pairing, unless $q_a=0$. This case will be considered separately. Setting $U=D=S=0$ and $Q=-X$ violates the anomaly cancellation conditions
unless $q_b=0$. This case will also be considered separately.

One may preserve the entire flavor group by assuming $Q=X=0$. We then get two non-abelian gauge groups $SU(M)$ and $G$ that
can become strong without disturbing each other. They can both be chiral prior to Higgs symmetry breaking without having anomalies.
This does not contradict the result of section \ref{SingleStack}, because the cubic and trace anomalies do not have to cancel within one stack, but
between the two stacks. But in that case $SU(M)$ necessarily couples to a chiral $U(1)$ and will break it. There is no flavor
group to take into account. 

If $X \not =-Q$, and in the absence of chiral rank-2 tensors,
complete flavor symmetry breaking can be avoided if we extend the flavor group with the $U(1)$ factor $Y$, because then its representation in both vectors and
anti-vectors has the same dimension. Now it is at least possible in principle to obtain a solution.
It follows that any solution must involve the $U(1)$, and that there can be only one $U(1)$, because
otherwise a linear combination would live entirely within $G$, which was already ruled out. 

\vskip .3truecm
\leftline{\bf The Main Argument}
\vskip .2truecm

We will now determine the possibilities for the surviving $U(1)$, assuming $Q\not=-X$.
It will in any case be a linear combination of a generator of the 
non-abelian flavor group $SU(N)$ and $Y$. 
\begin{equation}
\label{GeneralQ}
Q_{\rm em} = \Lambda+Y
\end{equation}
Note that we use the entire unbroken flavor group  here. The Higgs just breaks
$SU(N)$ to a subgroup, which by dynamical symmetry breaking is broken to a smaller subgroup. But in any case, the final result is 
of the form (\ref{GeneralQ}), with $\Lambda={\rm diag}(\lambda_1,\ldots,\lambda_N)$, and $\sum_i \lambda_i=0$.
The advantage of working with the full group $SU(N)$ is that the results can be applied directly to all choices for the broken
subgroup $G$ listed above. Furthermore it will contain all possibilities of dynamical symmetry breaking of the flavor group as well
as the most general Higgsless case.

To avoid massless charged leptons it is in any case essential to avoid chiral ones. This implies that 
the trace of $Q_{\rm em}$ in the lepton sector much vanish. Note that this trace is also the leptonic contribution to
the mixed anomaly of $Q_{\rm em}$ with gravity. So if this trace does not vanish, the strong $SU(M)$ interactions would have to produce chiral
massless charged baryons to match it. But we have already assumed that this will not happen.
This trace yields the equation
\begin{equation}
\label{LeptonTrace}
-L+(N-1)E+(N+1)T=0 \ .
\end{equation}
which can be added to the set of anomaly equations. 

These can now be solved completely in terms of $C_1$ and $C_2$. The result is
\begin{eqnarray*}
U &=& \tfrac{3+M}6C_1\\
S  &=& \tfrac{3-M}6C_1\\
D &=& NC_2 -\tfrac{M}{3} C_1\\
L\tilde q_b&=& -NC_2 +\tfrac{M}3 C_1\\
E\tilde q_b&=&-\tfrac12 C_2+\tfrac{M}{6} C_1\\
T\tilde q_b&=&-\tfrac12 C_2-\tfrac{M}{6} C_1
\end{eqnarray*}
For $M=3$ this implies $S=0$. As was the case for $N=2$ this follows from lepton charge pairing, but under the slightly stronger
condition that no non-abelian factor is left unbroken in the flavor group. Note that in the case $N=2$ we also used $T=0$. We see
now that only (\ref{LeptonTrace}) is needed. 

The derivation of (\ref{LeptonTrace}) holds only if there is a non-vanishing contribution to $Q_{\rm em}$ from $Y$. Hence it can be avoided 
if the Higgs mechanism breaks the $Y$ charge. This happens for the $SU(N) \rightarrow SO(N)$ breaking pattern for $H={\bf T}$, and
for the $SU(N) \rightarrow Sp(N)$ breaking pattern for $H={\bf E}$,  $N$ even. We will discuss these cases separately in 
section \ref{Exceptional}.

Now consider charge pairing in the strong interaction sector.
The charged $SU(M)$ vectors from $Q$ and $X$ can be paired with $D$, and with $U$ if $M=3$ (we only consider $M \geq 3$ here).
But in order not to have to distinguish separate cases we allow pairing  with $U$  for any $M$. We even leave $S\not=0$ and
allow $SU(M)$ vectors to pair off with these tensors. This can only happen for $M=1$, but there is no harm in
allowing $S\not=0$ for all other values of $M$.
The charges of the field $Q$ are of the form $q_a+q_b+\lambda_i$, and  those of $X$ are $q_a-q_b-\lambda_i$.
The charges $Q_{em}$ can have the special values $\pm q_a$ and $\pm 2q_a$. For all other  values cancellation with 
$U$ and $S$ is not possible. 

We will require more than just charge pairing, but also that the full $SU(M)\times U(1)_{\rm em}$ representation is non-chiral.
Consider a value for $\lambda_i$ so that $q_b+\lambda_i=\alpha q_a$. 
For $\alpha=0$ both {\bf Q} and {\bf X} contribute
vectors $(V,q_a)$ that can pair with each other and with {\bf D}. For $\alpha=-3$, {\bf Q} yields vectors $(V,-2q_a)$ which can be
paired with $U$, and {\bf X} yields vectors $(V,4q_a)$. For $\alpha=3$ the result is the same 
with $Q$ and $X$ interchanged. The vectors $(V,4q_a)$ can then be paired between $X$ and $Q$.
This means that the values of $\alpha$ that allow pairing with either {\bf D}, {\bf U} and {\bf S}
are constrained to $0$ and $\pm3$. 

Consider first any value of $\alpha$ other than $0, \pm3$. 
Suppose $\Lambda$ contains $n_i$ such eigenvalues. Then {\bf Q} contributes $Qn_i$ representations
$(V,(1+\alpha)q_a)$ and {\bf X} contributes $n_iX$ eigenvalues $(V,(1-\alpha)q_a)$. We must introduce $n_j$ additional charges 
$q_b+\lambda_j=-\alpha q_a$ to cancel these. Then the conditions for pairing are
\begin{eqnarray*}
n_i Q+n_j X&=&0\\
n_j Q+n_i X&=&0
\end{eqnarray*}
which implies $(n_i+n_j)(Q+X)=0$ and $(n_i-n_j)(Q-X)=0$. Since $Q+X\not=0$ we find that $n_i+n_j=0$, and since both
are positive integers, both must vanish. Hence charge pairing only allows $\alpha = 0, \pm3$.

Now we write down the pairing conditions for these values. We denote the number of eigenvalues with $\alpha=0$ as $n$, those
with $\alpha=-3$ as $n_Q$ and those with $\alpha=+3$ as $n_X$. We get the following pairing conditions for  the charges
$q_a$, $-2q_a$ and $4q_a$, respectively
\begin{eqnarray*}
n (Q+X)&=&D\\
n_Q Q+n_XX&=&U+S=-\rho(Q-X)\\
n_XQ+n_Q X&=&0
\end{eqnarray*}
Here $\rho\equiv \tilde q_b$, and in the second line the first anomaly condition (\ref{Aneq}) was used. The last two equations imply
$(n_Q+n_X)(Q+X)=-\rho(Q-X)$. Furthermore we we can solve them for $n_Q$ and $n_X$:
\begin{eqnarray*}
n_Q=-\frac{Q\rho}{Q+X}\\
n_X =\frac{X\rho}{Q+X}
\end{eqnarray*}
Now we must require that the total number of eigenvalues is equal to $N$, and that ${\rm Tr}\ \Lambda=0$.
This leads to the conditions
\begin{eqnarray*}
N&=&n+n_Q+n_X=\frac{D-(Q-X)\rho}{Q+X}\\
0&=&n(-q_b)+n_Q(-q_b-3 q_a)+n_X(-q_b+3)=-Nq_b+3(n_X-n_Q)q_a\\ 
&=&-Nq_b + 3\rho q_a=\tilde q_a \rho (1-\frac3{M})
\end{eqnarray*}
and hence $\rho=n_X-n_Q$.
The first condition implies $D\tilde q_a=NC_2-C_1$. Combining this with the first and third anomaly conditions (\ref{Aneq})
we find that $S\tilde q_a=(3-M) C_1$. Hence we find that quark charge pairing, even under the general conditions we 
allow, is possible only if $M=3$ and $S=0$. 

The lepton charges are equal to integer multiples of $\lambda_i+q_b$, where  $\lambda_i$ can have the values 
$-q_b + \alpha q_a$, with $\alpha=0, \pm3$. Hence all lepton charges are integer multiples of $3q_a$. Since $M=3$ and $S=0$ 
we have standard QCD with color singlet bound states whose charges are multiples of $3q_a$ as well. 
Hence the entire class of solutions
satisfies the same quantization rules of the Standard Model. This is a consequence of the fact that matter that only couples to the $U(3)$ brane can have charges $2q_a$ and $-q_a$,
that automatically satisfies the quantization rule. Then the  pairing requirement forces all  matter from open strings ending on both branes to satisfy the same quantization
rule. The generality of this result is remarkable. It holds for 
$SU(3)\times SU(N)\times U(1)$ for arbitrary $N$, with the flavor group broken
by any Higgs one may wish to consider, and with on top of that any dynamical symmetry breaking that could plausibly occur.
This has nothing to do with GUTs. Indeed, for arbitrary $N$  the gauge group does not even fit naturally in a GUT group.
   
We will now derive the multiplicities of all other fields. We have seen above that
\begin{equation}
D=n(Q+X)
\end{equation}
Since $S=0$ the first anomaly equation (\ref{Aneq}) implies
\begin{equation}
U=-\rho(Q-X)=(n_Q+n_X)(Q+X)=(N-n)(Q+X)
\end{equation}
Now we can use the remaining anomaly equations to determine $L$, $E$ and $T$:
\begin{eqnarray}
\label{LeptonMult}
L &=& nR\nonumber\\
E &=& \tfrac12 (N-n+1)R\\
T &=& -\tfrac12 (N-n-1) R\nonumber
\end{eqnarray}
where $R$ is the ratio
\begin{equation}
R=-\frac{Q+X}{\rho}
\end{equation}
which must be integer, since it is equal to $E+T$.

We now adopt the canonical Standard Model charge normalization, $q_a=-\tfrac13$. An $SU(N)\times U(1)$ vector
$(V,q_b)$ (the conjugate of {\bf L}) has the following decomposition 
 into charge eigenvalues:
\begin{equation}
(V,q_b)  \rightarrow  n \times \{0\} +  n_Q \times \{1\} + n_X \times \{-1\}
\end{equation}
%
%
%
Now we must check that 
the lepton sector is non-chiral if we choose one of these $U(1)$'s.  
The decomposed spectrum from {\bf L}, {\bf E} and {\bf T} contains charges $-2,-1,1,2$. So we get two equations, one for the charges
$\pm1$, and one for the charges $\pm 2$. These equations are:  
\begin{eqnarray*}
\left[-L  + (T  + E)n \right] &=& 0 \\
 \left[T (N-n+1) +  E  (N-n-1)\right]  &=& 0 
\end{eqnarray*}
Substituting the relations for $L, T$ and $E$ given above, we see that these conditions are satisfied.

Although charge pairing in the lepton sector is satisfied, this does not mean that this answer is acceptable. 
The charged leptons cannot get a mass from dynamical symmetry breaking, and hence must get it from the Higgs.
Here the single Higgs hypothesis plays an important r\^ole. If we were to allow more than one Higgs, one could break 
the flavor group down to just $U(1)$'s by means of a sequence of non-aligned vector Higgses, giving mass to all leptons.
But a single Higgs cannot
do that. The flavor group is eventually broken by the strong interactions which cannot generate a lepton mass. 

The simplest constraint on massless charged leptons is due to fermions in the representation {\bf T}. This representation
cannot have a Higgs coupling with {\bf E} or itself. The required Higgses must have a charge $0$ or $-4q_b$, and these
representations are simply not available. Fermions {\bf T} can couple to fermions {\bf L}, but then the Higgs must be in the
representation {\bf L} and hence
the breaking pattern is (\ref{VectorHiggs}). We have to determine which components of {\bf T} acquire a mass from this coupling.
This can be done by explicitly writing the Higgs field in vector or matrix form
and examining the resulting mass matrices. However, a much more efficient method is to observe that these computations do not
depend on $N$, and consider large $N$. 
If a component of a field is chiral with respect to the broken gauge group and has no counterpart 
in the field it is coupled to by the Higgs, then it cannot
get a mass. Then this must be true for all $N$.
We find that
the first component of {\bf T}, the anti-symmetric tensor of $SU(N-1)$, cannot acquire
a mass from this coupling, because it has no counterpart in  {\bf L}. Any non-zero
eigenvalue $\lambda$ gives rise to massless fermions with charge $2\lambda$ from {\bf T}. 
This can only be avoided if all $N-1$ eigenvalues vanish, hence
$n=N-1$, and we see from (\ref{LeptonMult}) that $T=0$, for any Higgs choice.

Now consider the Higgs choice $H={\bf E}$ and breaking pattern (\ref{AsymTensorHiggs}).
The only Yukawa couplings that are allowed are ${\bf LL}H$. 
This can only give mass to the $(1,2,0)$ component of {\bf  L}, since the other components are chiral.  
To avoid massless charged fermions, we would have to require that $SU(N-2)\times U(1)$ breaks down
to just  charges   zero, but that is not possible because the overall $U(1)$ charge of this multiplet does
not vanish, and hence ${\rm Tr}\ Q_{\rm em} \not=0$ on this component. 
If we use the breaking pattern (\ref{Symplectic}), the ${\bf LL}H$ Yukawa coupling can only give mass to the
first component $(1,V,0)$ of {\bf L}, which is not chiral since $V$ is a vector of $Sp(N-1)$. Then
the second component is a massless charged singlet independently of any further dynamical symmetry
breaking. Hence if the Higgs is {\bf E} we have to set $L=0$. This would imply $n=0$, and hence $N=1$, in which
case {\bf E} does not even exist.

This leaves the breaking pattern (\ref{VectorHiggs}) available for $H={\bf L}$ and $H={\bf T}$. Since $n=N-1$ we
see from (\ref{LeptonMult}) that $E=-(Q+X)/\rho$, which is always non-zero. If  $H={\bf T}$ there is no allowed 
Yukawa coupling to leptons at all, so let us consider $H={\bf L}$. In this case the first component of 
fermion multiplets {\bf E} remains massless. The only way to avoid massless charged leptons from these
components is if the eigenvalues of $Q_{\rm em}$ on the $N-1$ dimensional subspace are all zero,
or if the subspace is one-dimensional, so that the anti-symmetric tensor component does not exist.
The first option is not available, because the trace over the subspace is $-Nq_b$. Hence we find $N=2$.
There is no need to investigate this case any further, because this is the Standard Model. For $M \geq 3$
and $N \geq 1$ this is the only solution to our conditions (see section \ref{UthreeUone}), apart from cases
with uncontrollable strong interactions.

\subsubsection{Examples}

If $H={\bf L}$ the simplest possibility is that $SU(N)$ breaks to $SU(N-1)\times U(1)$ and then $SU(N-1)$
is broken completely by the strong interactions. This means that $\Lambda=0$ within $SU(N-1)$, hence
$n=N-1$, and $n_Q=1, n_X=0$ (or vice-versa). Hence $\rho=-1$ and $X=0$. This results in a generalization of the
Standard Model with $U=E=Q, D=L=(N-1)Q$ and $T=S=0$. A family has the form
\begin{equation}
\label{GenSM}
\left(3,N,-\tfrac13+\tfrac{1}{N}\right)+\left(\overline{3},1,\tfrac13\right)+(N\!-\!1)\left(\overline{3},1,-\tfrac23\right)+(N\!-\!1)\left(1,N,-\tfrac{1}{N}\right)+\left(1,A,\tfrac{2}{N}\right)
\end{equation}
Note that all $Y$ charges satisfy the rule
\begin{equation}
Y=-\frac{t}3 + \frac{s}{N} \ \hbox{mod}\ \ 1
\end{equation}
where $t$ is $SU(3)$ triality and $s$ is $SU(N)$ N-ality. This is the generalization of the Standard Model result to arbitrary $N$, and
implies charge integrality of color singlets. But we have already seen that the latter statement is valid even more generally for all
Higgs choices, combined with any allowed chiral symmetry breaking.

The other breaking patterns that are available for given $N$ require different choices of particle multiplicities
$Q,U,D,\ldots$, so that there is at most one pattern available in each case. Whether they can be realized or not depends
on strong dynamics, and all we can say is that they are not forbidden. 

If $N=2$ there are two possible patterns (since $n_Q\not=n_X$), namely 
$n_Q=1,n_X=0,n=1$ and $n_Q=2, n_X=n=0$. The first case correspond to the Standard Model, and can be obtained using the
Higgs {\bf L} and in the Higgsless case by means of dynamical symmetry breaking. The second case gives $\rho=-2$, $X=D=L=0$, 
$U=2Q$, $E=\frac34 Q$ and $T=\frac14 Q$. Hence $Q$ must be a multiple of 4. A family has the form
\begin{equation}
4 (3,2,\tfrac23)+8(\overline{3},1,-\tfrac23)+(1,3,-2)+3(1,1,2)
\end{equation}
The $SU(2)$ factor would be broken completely by QCD, leaving a non-chiral spectrum. But as we have already seen in the general analysis
above, this
spectrum is plagued by massless charged leptons.

\subsubsection{Asymptotic Freedom}

In example (\ref{GenSM})
the number of color triplets is a factor $N/2$ larger than
in the Standard Model, and hence for $N > 16$ asymptotic freedom of $SU(3)$ is lost. This means that our assumption 
that $SU(3)$ is the strongest interaction at low energies is violated. It may already be violated for smaller values of $N$ if the coupling constant
of the second brane stack is large enough.
One could consider the possibility that  a strong non-abelian
factor $G$ left over from $SU(N)$ takes over the r\^ole of $SU(M)$.  This is problematic, since $G$ becomes part of a larger group
$SU(N)$ at the Higgs scale, and hence $G$-baryon number may be violated by weak interactions. Then $G$ would be unsuitable
for making stable hadrons. 
In the next section (just after Eq. (\ref{LikeSM})) we will
discuss an example where that is indeed what happens. Perhaps this can be turned into a general argument against the use
of using such groups -- remnants of Higgs breaking -- as strong interaction groups, but we have not pursued this
possibility. In this paper we will simply put such spectra in the ``undecidable" category.

\section{Special Cases}\label{Exceptional}

Here we discuss special cases not covered before: the possibility that the electromagnetic charge
does not contain a contribution from $Y$ at all, or that $Y$ is embedded in the phase factor
of just one of the two unitary groups. We also consider the case $M < 3$.

\subsection{Breaking of $Y$ by rank-2 tensors}\label{Ybreaking}

The symmetric and anti-symmetric tensor Higgses can break $SU(N)\times U(1) \rightarrow  SO(N)$ for $H={\bf T}$ or 
$SU(N) \times U(1) \rightarrow  Sp(N)$ for $H={\bf E}$, $N$ even (for $N$ odd see the previous
section). In general there is then no $U(1)_{\rm em}$ left over after Higgs breaking. Only if 
$H={\bf T}$ and $N=2$ there is still an $SO(2)$ subgroup that could play this r\^ole.
In all other cases, there is only a chance of meeting condition 5a if the strong interactions
somehow break $SO(N)$ or $Sp(N)$ to a $U(1)$.
Since $Y$ is broken, we cannot use the lepton charge pairing condition (\ref{LeptonTrace}), which
led to the conclusion that $S=0$ for $M=3$. Therefore analyzing this case requires stronger
assumptions about dynamical symmetry breaking. Note that $Y-$breaking by $H={\bf T}$ for $M=1$ and $M=2$ is 
discussed below in section  \ref{TensorHiggs12}.

If we impose the condition $S=0$ (and also $U=0$ for $M>3$) we get a spectrum consisting only
of $SU(M)$ vectors and anti-vectors, of the form $Q(V,V)+X(\overline{V},V)+D(\overline{V},1)+U(\overline{V},1)$. 
As in the previous section, we can ask which subgroup of the $SO(N)$ or $Sp(N)$ flavor gauge group we can choose 
so that this spectrum is non-chiral. For arbitrary $Q$ and $X$ this has no solution. Only if $Q=-X$ such a subgroup
exists, and then it can be any subgroup, including $SO(N)$ or $Sp(N)$ themselves.
To pair off the remaining quarks we must
have $U=-D$ (if $M=3$) and $U=D=0$ (if $M > 3$).
For the anomaly conditions (\ref{Aneq}) this implies $C_2=0$ and $C_1=-2Q\rho$, where we have set
$\tilde q_a=1$ and $\rho=\tilde q_b/\tilde q_a$ as before. Then we read off that 
$U=-2Q\rho$, $E=-T=Q\rho$, $D=2(4-M)Q\rho$, $L=2(4-M)Q$. We observe that pairing
of {\bf U} and {\bf D}, {\it i.e.} $U+D=0$ requires $M=3$. Hence for $M > 3$ there is no solution,
but for $M=3$ there is: $X=-Q$, $S=0$, $D=-U=2Q\rho$, $E=-T=Q\rho$ and $L=2Q$. This solution
is valid for any rational value of $\rho$ provided $Q$ is chosen so that $Q\rho$ is integer. 

We see that requiring non-chirality in the quark sector leads to $M=3$ and both $E$ and $T$
non-zero. Consider now the Higgs coupling in the lepton sector.
The only allowed 
leptonic Yukawa couplings are, for $H={\bf T}$
\begin{equation}
H{\bf L}{\bf L} \ \ \hbox{and}\ \  H{\bf T}{\bf N}\ ,
\end{equation}
and for $H={\bf E}$
\begin{equation}
H{\bf L}{\bf L} \ \ \hbox{and}\ \  H{\bf E}{\bf N}\ ,
\end{equation}
where {\bf N} is a neutral particle, like  a right-handed neutrino in the Standard Model. We leave its origin unspecified,  but clearly
it cannot come from open strings ending on one of the branes. The second terms only give a mass to
the $SO(N)$ or $Sp(N)$ singlet that must always exist in the decomposition of {\bf T} or {\bf E}
to $SO(N)$ or $Sp(N)$ respectively; the singlet represents  the vacuum direction of the Higgs field. Like in the Standard Model,
{\bf N} does not have to be present, and then the singlet components of {\bf T} or {\bf E} do not get a mass. 
In the first terms ${\bf L} {\bf L}$ 
must be symmetric in the full flavor group $SU(LN)$, {\it i.e} the
group acting on all components of $L$ multiplets {\bf L}.
This implies that for $H={\bf T}$ the coupling must be symmetric in the family index, whereas
for $H={\bf E}$ it must be anti-symmetric. Then for $L=1$ and $H={\bf E}$ this coupling does not exist. 

For $N=2$ and $H={\bf T}$ this implies that there are massless charged leptons in the spectrum, because only
the neutral entry of ${\bf T}$ acquires a mass, and $T=-Q\rho\not=0$. For $ N > 2$ or $N=2, H={\bf E}$ we have no such problem, because
we do not even have a $U(1)$. But  if $SO(N)$ or $Sp(N)$ break to a $U(1)$ we will have
almost certainly
the same problem. The only way out would be that $SO(N)$ or $Sp(N)$ break to a group
$G\times U(1)$, and that all leptons are in non-trivial representations of $G$, so that their
masslessness is not immediately fatal. Then we would have to consider the dynamics of $G$ to
check if it breaks $U(1)_{\rm em}$. 

In the previous section this was circumvented by determining which subgroups of the flavor group could
give a real representation, but here the flavor group is already (pseudo)-real, so this method will not give 
further constraints. In fact, there is no reason why the flavor group should be broken dynamically, because
all quarks can get a dynamical mass without breaking it.

Because the quarks have a rather simple coupling to the flavor group, we can say a bit more. The strong interaction
condensate will be in the representation $V \otimes V$ in any of the  non-abelian factors of $G$ and is uncharged under $Y$.
We can now use something analogous to the most attractive channel (MAC) hypothesis of \cite{Raby:1979my} to decide which breaking
pattern is the most likely one.

The MAC hypothesis was originally proposed to study self-breaking of gauge theories used in tumbling gauge theory 
scenarios. Here we do not need it for that purpose, because for a vector and an anti-vector of $SU(M)$ the most attractive
channel is an $SU(M)$ singlet, which does not break $SU(M)$. But in addition to the $SU(M)$ potential there will be a small
single vector exchange potential due to all flavor gaugings, which will determine which direction is chosen in flavor space. This potential is of the
same form as the one proposed in \cite{Raby:1979my}, namely
\begin{equation}
V \propto \frac{g_b^2}{r}\left[(\Delta_c-\Delta_1-\Delta_2\right]  \ ,
\end{equation}
where $\Delta_c$, $\Delta_1$ and $\Delta_2$ are the quadratic Casimirs of the condensate and the quark and anti-quark it is made of. 
The most attractive channel is the one that minimizes $V$, {\it i.e.} the smallest Casimir in the product $V\otimes V$. 
For orthogonal and symplectic groups this tensor product  always contains a singlet, which of course has the smallest $\Delta_c$.
This suggests that these groups are not broken at all, and in particular that no electromagnetic $U(1)$ is produced.

Hence we conclude that all of these possibilities are ruled out if we assume $S=0$. If $S\not=0$ there is no obvious solution
for the strong interaction spectrum, and it is possible that the strong interaction group $SU(M)$ itself has to be broken. We regard this case as undecided. 
The low energy spectrum is obviously chiral, so this violates condition 5b.

\subsection{$q_a=0$}

If $q_a=0$ the solution of the anomaly conditions (\ref{Aneq}) is 
$X=Q$, $L=T=E=0$, with $U, S, D, Q$ and $X$ subject to the $SU(M)$ anomaly cancellation condition. 
The resulting unbroken $SU(M)\times SU(N)\times U(1)_{Y}$ spectrum is
\begin{equation*}
Q[(V,V,1)+(V,\overline{V},-1)]+\hbox{flavor-neutral {\bf U}, {\bf D}, {\bf S} matter} \ , 
\end{equation*}
with the flavor-neutral matter canceling the $SU(M)$ anomaly. 
Since only $SU(M)$ vectors (and no anti-vectors)  couple to the flavor gauge group, there is no combination of an $SU(N\!-\!1)$ generator and the $U(1)$ that
is non-chiral with respect to $SU(M)$. Hence for $M\geq 3$ the most plausible assumption is therefore that there will not be electromagnetism.
There are solutions with $Q=0$, but then one only gets an $SU(M)$ gauge group, and no electromagnetism.
For $M=1$ and $M=2$ the entire spectrum is  non-chiral before symmetry breaking, so that assumption 3 is violated.

\subsection{$q_b=0$}\label{QbIsZero}

If $q_b=0$, the anomaly cancellation conditions imply that $Q=-X$ and $U=S=D=0$. Then
the matter representation before symmetry breaking is
\begin{equation*}
Q[(V,V,1)+(\overline{V},V,-1)]+\hbox{$Y$-neutral {\bf L}, {\bf E}, {\bf T} matter}
\end{equation*}
For $N=1$ and $N=2$ this case can be discarded since the high energy theory is non-chiral. For $N \geq 3$ it is chiral. Since we are
assuming throughout this paper that $SU(M)$ remains unbroken, the only candidate Higgses are uncharged, and cannot break $Y$. 
Furthermore $Y$ is non-chiral with respect to $SU(M)$, so there is no reason why it should break dynamically. 

The three Higgs choices {\bf L}, {\bf T} and {\bf E} break $SU(N)$ to some group $G$. If $G$ has only real representations, then
we have found a solution to our conditions. This happens if $H={\bf L}$ or $H={\bf T}$ and $N=3$
(with $SU(3)$ breaking to $SU(2)$), 
if $H={\bf T}$  for all $N$ 
if we choose the breaking to $SO(N)$, and $H={\bf E}$ for $N=4$ (breaking $SU(4)$ to $SU(2)\times SU(2)$)
and for all $N$ if we choose the breaking to $Sp(N)$ or $Sp(N-1)$.
We do not have to worry about massless charged free leptons, because there are no charged free leptons at all. Hence this case provides a solution to all our conditions, in the form 5a or 5b,  for all $M$ and all $N \geq 3$.  

Since the presence of free leptons was not part of our requirements we do not discard these cases. To get some sort of atomic physics,
strong interaction bound states with opposite charges must somehow make atoms. 
The strong interactions will in general break the gauge group, and one can make a plausible guess about how it is broken. Obviously,
$G$ (which from now on can be any of the non-abelian factors obtained from the Higgs mechanism, or $SU(N)$ itself) must
break to a subgroup that has real representations. Is it possible that this subgroup contains an additional $U(1)$ factor, so that
the fermions {\bf L}, {\bf T} and {\bf E} could produce charge free leptons after all?

Once again we can use the MAC hypothesis explained above.
This suggests that orthogonal and symplectic groups are not
broken at all. For $SU(N)$, $N>2$, the most attractive channel is the anti-symmetric tensor. This may break $SU(N)$ to a symplectic 
group or $SU(N-2)\times SU(2)$, but in neither case there is an additional $U(1)$ factor. In fact, the only way one might have
obtained an additional $U(1)$ is from the breaking of $SO(N)$ or $Sp(N)$ by rank-2 tensors (for example $SO(2N) \rightarrow SU(N)\times U(1)$), but we have just seen that the rank two tensors are a less attractive channel than the singlet. Therefore it is not likely that
there will be an additional $U(1)$, and $U(1)_Y$ will have to play the r\^ole of electromagnetism.

Condition 5c can also be satisfied in some cases, so that
 all fermions can get a mass from the Higgs mechanism. We will only
discuss the case $H={\bf T}$ to demonstrate this. This Higgs has  couplings 
$H^*{\bf Q}{\bf X}$ and $H{\bf L}{\bf L}$ that can give mass to all components of {\bf Q}, {\bf X} and {\bf L}.
The fermionic fields {\bf T} and {\bf E} are not needed to cancel the $SU(N)$ anomaly, only {\bf L} is needed.
Its multiplicity must then be $L=2MQ$. Then we end up with a low energy $SU(M) \times SO(N) \times U(1)$ spectrum 
\begin{equation}
Q[(V,V,1)+(\overline{V},V,-1)+2 M (1,V,0)]
\end{equation}
This is a solution to all our conditions in the form 5a, 5b and 5c. 
It is chiral at high energies, and the Higgs renders it non-chiral and gives mass
to all fundamental fermions. 

However, there is an obvious problem. Let us assume that $M$ is odd, because for $M$ even there would be no
fermions in the spectrum at all, and the prospects look worse.
Quarks have positive electric  charges, anti-quarks negative ones. This means
that baryogenesis, starting from zero baryon number  and zero charge cannot work because of electric charge
conservation. Furthermore, even if one could somehow make an asymmetric universe, for example by starting with
an asymmetric initial state and not having inflation, then all charged particles have positive charges. There would
be no negative charges and no atomic physics. 

There may still be a way out. Perhaps a positively charged quark from the second family could be used
together with a negatively charged anti-quark from the first family. They could make baryons of charge $M$ and
anti-baryons of charge $-M$. To prevent these from annihilating each other or decaying into first family quarks, 
all interfamily transitions must be forbidden, {\it i.e.} the CKM matrix of the weak interactions must be strictly
diagonal. Perhaps a discrete symmetry can be invented to achieve that.
Then the charge $M$ baryons could be given a much larger mass than the charge $-M$ anti-baryons 
by creating a large mass hierarchy between the first and second family by means of the Yukawa couplings.
 The charge $-M$ baryons could then act as 
leptons. To make this work one would  have to invent a baryogenesis-like process that greatly enhances the
abundance of positively charged second family quarks and negatively charged first family anti-quarks with
respect to their anti-particles. 

Even if this implausible scenario can be realized, there are many other problems.
The baryon spectrum has only charged particles, no ``neutrons". Nuclear physics like we observe is not
possible because of electromagnetic repulsion. Furthermore there is a leftover ``weak" non-abelian gauge group
with unknown implications.

We cannot rule out this scenario completely, but this discussion makes us appreciate even more how nicely all of these issues
are solved in the Standard Model: there are stable baryons with positive and zero charge, and leptons with negative
charge; baryogenesis is not forbidden by electric charge conservation and there is a  full CKM matrix without spurious zeroes.

\subsection{$q_a=q_b=0$}\label{qaisqbiszero}

It may happen that both $U(1)$'s are broken due to axion mixing in string theory. Then the anomaly cancellation conditions
reduce to just the non-abelian anomalies of $SU(M)$ and $SU(N)$.  The Higgs mechanism and/or chiral symmetry breaking
will have to produce $U(1)_{\rm em}$ as a subgroup of $SU(N)$.

The only way a $U(1)$ can emerge directly from the Higgs mechanism is via the breaking of $SU(2)\rightarrow SO(2)$ 
by $H={\bf T}$ (as discussed in  section \ref{HiggsChoice}, adjoint breaking may produce a $U(1)$, but will not turn a chiral
theory into a non-chiral one).
If there is no $Y$ charge, all leptons are non-chiral and hence have a large mass by assumption 3. 
Note that there is some tension here between assumption 3 and non-abelian anomaly cancellation for $SU(2)$ as required
by string tadpole cancellation. Some of the non-chiral leptons {\bf L}, {\bf T} or {\bf E} may in fact cancel the $SU(2)$ anomalies of {\bf Q} and {\bf X}, and hence
one may question if they would really acquire a large (string scale) mass. We are assuming that field-theory arguments prevail, and that
any mass not forbidden by field-theoretic chirality will be generated, and will be large. Fortunately in the vast majority of cases the 
$SU(2)$ string-theoretic anomaly condition is redundant, so that this issue does not arise.

As in previous cases, we will only discuss dynamical symmetry breaking for $S=0$ ($M=3$) or $S=U=0$ ($M>3$), and consider
the other options as undecided. Also as before, the only chance for any subgroup of the flavor group to be left unbroken is if
$X=-Q$, and because of $SU(M)$ anomaly cancellation $U=-D$ for $M=3$ and $U=D=0$ for $M > 3$, because $U$ and $D$ cannot
be paired.  If $N=2$ and $X=-Q$, $U=-D$ the entire spectrum is non-chiral, and hence massive by assumption 3. 

In all other cases the analysis of the previous subsection applies, an we would not expect that the flavor group does not break to a $U(1)$.

\subsection{$M=2$}\label{SectionMTwo}

If $M < 3$ the foregoing analysis does not necessarily apply because
of $SU(M)$ $M$-dependent conjugation properties of anti-symmetric tensors $([M-r] \sim \overline{[r]})$
or over-saturation of  the bound $r \leq M$.
If $M=1$ anti-symmetric tensors do not exist,
and vectors, conjugate vectors and symmetric tensors are all equivalent to the identity. If $M=2$
the anti-symmetric tensor is the identity, and vectors and conjugate vectors are equivalent. If $M=3$
the anti-symmetric tensor is an anti-vector. These special features allow pairings that are
not possible for other values of $M$. Note that the cases $q_a=0$ and/or $q_b = 0$  have already been discussed above
for all $M$ and $N$, so here we assume that $q_a\not=0$ and $q_b\not=0$.

We start with $M=2$. Then
  the lepton trace condition (\ref{LeptonTrace}) is not valid, because the representation {\bf U} is
an $SU(2)$ singlet, and hence a lepton as well. Instead of (\ref{LeptonTrace})  we get then
\begin{equation}
U\tilde q_a+\left[(N-1) E+(N+1) T-L\right]\tilde q_b=0
\end{equation}
If we add this to the anomaly equations (for any $M$), we find the following two-parameter solution
\begin{eqnarray}
\label{SUTWO}
S&=&\tfrac15 (2-M) C_1 \nonumber \\
U&=&\tfrac15 (3+M) C_1  \nonumber \\
D&=&-\tfrac15(3M+4) C_1 + N C_2\nonumber \\
L \tilde q_b&=&\tfrac15(3M+4) C_1 -N C_2\\
E \tilde q_b &=&-\tfrac1{10}(2M+1) C_1 -\tfrac12 C_2 \nonumber \\
T \tilde q_b&=&\tfrac1{10}(2M+1) C_1 -\tfrac12 C_2 \nonumber
\end{eqnarray}
We see that just as happened before for $M=3$, this modified lepton trace condition eliminates the symmetric tensor {\bf S} also for $M=2$.
As before, this trace condition is valid only if $Y$ makes a contribution to $Q_{\rm em}$. If $Q_{\rm em}$ is embedded entirely
in the non-abelian $SU(N)$ group a separate argument must be made.

Another important consequence of (\ref{SUTWO}) is that $U$ never vanishes. The representation ${\bf U}$ does not couple
to the flavor group at all, and hence it is a charged free lepton, that must get a mass. It may either get a mass at the large
scale if it can be paired with another field, or it must be coupled to another field by the Higgs boson. The only other field it can be
paired with is {\bf E} for $N=2$. This requires $q_b=-q_a$ and $U=E$ (or the  equivalent solution with sign changes 
for $q_b$  and lepton multiplicities).
Conditions (\ref{SUTWO}) combined with the pairing condition 
 $U=E$ imply $C_1=C_2$, and hence $X=0$, $C_1=C_2=Q$.
Then the full spectrum is given by $S=0, U=Q, D=0, L=0, E=Q, T=0$. But since $q_b=-q_a$,  {\bf Q} has charge zero  and is non-chiral, so that
the entire spectrum is non-chiral before the Higgs mechanism. So this possibility is ruled out by assumption 3.

The other way {\bf U} can get a mass is by means of the Higgs mechanism. Consider a Higgs field $H$ in a
representation $(R,h)$, and a Yukawa coupling $H\psi\xi$, where $\psi$ is a field {\bf U} and
$\xi$ is the fermion it couples to (the coupling $H^*\psi\xi$ may also be considered, but leads to the same
conclusion).
Since ${\bf U}$ is an $SU(M)\times SU(N)$ singlet, $H$ can couple it to   
a field $\xi$ in the representation $(R^*,-h-2q_a)$. This is the conjugate of $(R,h+2q_a)$.
This representation does not exist for generic $N$, because
since $(R,h)$ exists, there is in general no representation $R$ with a different charge. 
The only possibility is then that $R$ is equivalent
to a different representation $\hat R$, which can happen in special cases for some small values of $N$. Then if $\hat R$ comes
with a $Y$-charge $\hat q$ it can couple to {\bf U} if $h+2q_a=\hat q$. 
The options are, for $H={\bf L}^*=(V,q_b), H={\bf T}=(S,2q_b)$ and ${H=\bf E}=(A,2q_b)$, respectively 
\begin{eqnarray*}
V&\sim &\overline{V}\ \ \ \ \  \hbox{for $N=2$ and $q_b+2q_a=-q_b$, with $U=L$} \\
V&\sim &\overline{A}\ \ \ \ \   \hbox{for $N=3$ and $q_b+2q_a=-2q_b$, with $U=E$}\\
S&\sim &\overline{S}\ \ \ \ \   \hbox{for $N=2$ and $2q_b+2q_a=-2q_b$, with $U=T$} \\
A&\sim &\overline{A}\ \ \ \ \   \hbox{for $N=4$ and $2q_b+2q_a=-2q_b$, with $U=E$}
\end{eqnarray*}
where the representation to the left of $\sim$ is $R$ and the one on the right is $\hat R$.
%

%
%
The first choice corresponds to standard breaking by a vector Higgs. We know that in that case ${\bf T}$
produces massless charged free leptons, and hence we must set $T=0$. But from (\ref{SUTWO}) we see that
for $N=2$, $L=4T$ so if $T$ vanishes, so does $L$. Then $U$ must also vanish. Hence both $C_1$ and $C_2$
vanish, and there is no solution. 

The second choice gives  $\tilde q_b=-1$ for a brane configuration $U(2)\times U(3)$ (we normalize $\tilde q_a$ to 1).
Imposing $U=E$  
implies $C_1=C_2$. Then we find $X=0$, $T=0$, $D=L=Q$, $U=E=Q$. This looks a
lot like the Standard Model. Indeed, it is just the Standard Model, but with $SU(3)$ as a ``weak" interaction group
broken by the color triplet Higgs, and $SU(2)$ as the strong interaction group. The color triplet Higgs {\bf L} (not to be confused
with the $SU(2)$ triplet Higgs {\bf T)} is the component of the $SU(5)$ Higgs boson that plays a r\^ole in  the infamous doublet-triplet splitting problem.
The Higgs {\bf L} can give mass to all fermions. We normalize the charge so that $H=({\overline V},2)$. Then $H$
couples to $SU(2)$ singlets forming Higgs multiplets of the form (\ref{SUTHREEmultiplet}) plus a half multiplet of doublets. One family 
has the following decomposition in terms of Higgs multiplets
\begin{equation}
\label{SpecialHM}
(1,{\cal H}(\overline{V},-4))={\bf U}+{\bf E}+{\bf L}+{\bf N}; \ \ \ \ \ \ (2,{\cal H}(V,1))={\bf Q}+{\bf D}
\end{equation}
The resulting low energy spectrum is,
in $SO(4)\times U(1)$ notation
\begin{equation}
\label{WrongSM}
(v,0)+(s,1)+(s,-1)+(c,1)+(c,-1)+(1,2)+(1,-2),
\end{equation}
where $v=(2,2)$, $s=(2,1)$ and $c=(1,2)$, and the representations are Weyl multiplets. The charged particles combine to Dirac multiplets,
whereas the $(v,0)$ particle has a Majorana mass. This means that a pair of such particles can disappear into the vacuum, analogous
to neutrino less double-beta decay, unless the Majorana mass is tuned to zero. 

One could treat either or both $SU(2)$ factors as a strong interaction. Such an interaction would be needed to create  a range of charged
nuclei, since the charges of the fundamental particles have too small a range to make anything interesting. However, there is no 
conserved baryon number, in contrast to the Standard Model. The second $SU(2)$ is involved in the weak interactions, which can turn doublets
into singlets, and the doublets of the first $SU(2)$ can annihilate into charged weak bosons. An analogous process in the Standard Model
is  $u{\bar d} \rightarrow W^{+}$, and of course this preserves baryon number. But if the strong interactions are $SU(2)$, such a process is capable of removing two
quarks from a bound state. The analog of baryon number is just a $\mathbf{Z}_2$ symmetry. The existence of stable hadrons
with large charges is therefore highly questionable.
However, respecting the principles
stated in the introduction we do not reject this example on the basis of additional anthropic criteria; we keep it on the list as ``acceptable".

The third option gives $\tilde q_a=1, \tilde q_b=-\frac12$.
The condition  $U=T$ implies $C_2=2 C_1$, {\it i.e.} $X=0$. Then $C_1=\frac12 Q$, $T=\frac12 Q$,
$D=Q$, $L=2Q$ and $E=\frac32 Q$. The minimal solution is for $Q=2$. The appearance of multiplets {\bf E} causes a problem,
because {\bf E} cannot couple to  a Higgs {\bf T}. For $N=2$, {\bf E} is a singlet and will remain in the spectrum as a charged, massless
free fermion. So this case is ruled out.      

The fourth option gives $\tilde q_a=1$, $\tilde q_b=-1$.
Hence $C_1=Q-X, C_2=Q+X$. Requiring $U=E$ yields $C_1=C_2$, so that $X=0$. 
The Higgs boson must be {\bf E}, which can break $SU(4)\times U(1)$ to $Sp(4)$ or $SU(2) \times SU(2) \times U(1)$.
In the former case $Y$ is broken, and then the foregoing arguments do not apply. In the second case, the field
{\bf Q} breaks up into components $(2,2,1,0)$ and $(2,1,2,\frac12)$. There is nothing else in the spectrum that can be paired
with the second component, and hence the result is not chiral.

This problem also occurs for $M=3$ and $N > 2$, and in that case we went a step further and allowed $SU(3)$ to break 
the flavor group. For $M=2$ this works in a different way. A theory with $F$ doublets of $SU(2)$ is believed to break
its $SU(F)$ flavor group to $Sp(F)$. Note that $F$ must be even because of the non-perturbative Witten anomaly \cite{Witten:1982fp}.
A bi-quark condensate has a wave function that is  anti-symmetric in spin and color, and hence it must be anti-symmetric
in flavor as well. This anti-symmetric tensor breaks $SU(F)$ to $Sp(F)$. But what we need to know is the fate of a gauge group
embedded as a subgroup of $SU(F)$. We will not attempt to work this out here.  
For $M=2$ we only list spectra that are made non-chiral by the Higgs mechanism, and then the fourth option is
ruled out.

\subsubsection{Symmetric Tensor Higgs}{\label{TensorHiggs12}

We now return to the case we postponed earlier.
If $Y$ is completely broken by the Higgs mechanism there is only one possibility for getting a massless $U(1)$ without
invoking dynamical symmetry breaking, and that is $H={\bf T}$, $N=2$, $SU(2) \rightarrow SO(2)$. In that case {\bf U} becomes
an uncharged massless field, which does not cause any problems. Fermions in the representation {\bf T} give rise
to massless charged free leptons, so $T$ must be zero. Fermions {\bf E} are uncharged, and fermions {\bf L} can all get a mass from
the Higgs through the $H{\bf L}{\bf L}$ coupling. The Higgs can  couple {\bf Q} to {\bf X}, but only generates masses for all doublets
 if $Q=-X$. If we impose
the conditions $T=0$ and $Q=-X$ we find that $C_2=0, C_1=-2Q\rho$, $E=0$, $U=S=-Q\rho$, $D=2MQ\rho$ and $L=2MQ$, with 
$\tilde q_a=1$ and $\tilde q_b=\rho$. We have written down the solution for any $M$.
For $M=2$ we get a strongly interacting $SU(2)$ theory which has doublets and triplets.
The only r\^ole of $U(1)_Y$ is to provide chirality to protect the fermions from getting a large mass. The broken spectrum consists
of $Q$ copies of $SU(2)\times U(1)$ families of the form
\begin{equation}
2(2,1)+2(2,-1)+(1,0)+4\rho (2,0)+(3,0)+4(1,1)+4(1,-1)
\end{equation}
There are $2Q$ doublets with charges $+1$ and $-1$. These are Weyl spinors, combining to $2Q$ Dirac spinors in the representation
$(2,1)$.
The doublets {\bf D} and the triplet {\bf S} cannot get a mass from the Higgs. There is no solution where all fundamental fermions get a mass.

The weak interactions are mediated by massive charge 2 vector bosons that couple the positive charges to the negative ones. Hence
if the quarks are heavier than the leptons, a bound state of two $(2,1)$ quarks can decay into two $(1,1)$ leptons. Depending on mass choices, mixings and binding energies, various scenarios are possible.  However in all models with $SU(2)$ strong interactions there is
no conserved baryon number (at best it is preserved modulo 2),  and it is hard to see how one can build up a spectrum with 
stable hadrons with an interesting range of charges. However, we will not explore this further, and put this model in the ``acceptable"
category. Conditions 5a and 5b are satisfied, but 5c is not.

The case $M=3$ was already discussed in section \ref{Ybreaking}, with the requirement $S=0$. Here we see that if we impose $T=0$ to avoid massless charged free leptons, we always get $S\not=0$.

%
%
%
%
%

\subsection{M=1}

The case $M=1$ is all that remains. Since we are requiring a strong $SU(M)$ gauge group, there is really no need
to consider this, but there are a few interesting solutions.
We will first analyze the symmetry breaking $SU(N) \rightarrow SU(N-1)\times U(1)$
caused by either $H={\bf L}$ or $H={\bf T}$.
We will only consider those spectra
that are non-chiral after Higgs symmetry breaking. We may assume $q_a\not=0$ and $q_b\not=0$, and 
$q_b < 0$.
All solutions  are listed in Table 1. We have also added as item nr. 8 the solution for $M=2$
discussed in the previous subsection, and a solution for $H={\bf E}$ discussed below in section \ref{AsymTensorHiggs}. 
The solutions obtained
by breaking $SU(2)$ to $SO(2)$ by a Higgs {\bf T} are included as item nr. 7, but note that they are non-chiral at low energies only for $M=1,2$.
Spectra where neither of the Higgs choices
{\bf L} or {\bf T} gives a mass to all charged free leptons have not been included. We also omitted one
case where the entire spectrum was already non-chiral before Higgs symmetry breaking, due to
a special identification that is only possible for $M=1$. 
 Finally, we have also included the Standard Model as item nr. 9. The table contains all cases, for all $M$ and $N$, where a Higgs breaks a chiral
 spectrum to a non-chiral one without massless charged leptons, with the exception of the class discussed in section \ref{QbIsZero}.
\begin{table}[h]
\begin{center}
\begin{tabular}{|l||c c||c c||c|c c r r r|rlc|r|} \hline
Nr. & $M$ &  $N$ & $q_a$ & $q_b$ & Higgs & $Q$  & $U$ & $D$ & $S$ & $X$ & $L$ & $E$ & $T$  \\ \hline
 1& $1$ &  $2$ &  $2$ &      $-3$     &    {\bf L}            & $3$  & $6$ & $3$ & $3$ & $0$ &   $1$ &   $1$ & $0$  \\ \hline
2& $1$ &  $2$ &  $4$ &      $-1$      & {\bf L}                        & $2$  & $1$ & $1$ & $0$ & $0$ &   $2$ &   $3$ & $1$  \\ \hline
3a& $1$ &  $2$ &  $2$ &      $-1$    & {\bf L}      & $3$  & $4$ & $1$ & $3$ & $-4$ &   $1$ &   $0$ & $-1$   \\ 
3b& $1$ &  $2$ &  $2$ &      $-1$    & {\bf L}      & $2$  & $2$ & $1$ & $1$ & $-1$ &   $1$ &   $1$ & $0$   \\ 
3c& $1$ &  $2$ &  $2$ &      $-1$    & {\bf L}      & $4$  & $5$ & $0$ & $3$ & $-4$ &   $0$ &   $1$ & $-1$   \\ \hline
4 & $1$ &  $3$ & $3$ &         $-2$  & {\bf L}    & $2$  & $3$ & $2$ & $1$ & $0$ & $1$ & $1$ & $0$   \\ \hline
5 & $1$ &  $3$ & $3$ &         $-1$  & {\bf E}    & $0$  & $0$ & $-2$ & $-1$ & $1$ & $-2$ & $1$ & $0$   \\ \hline
6& $1$ &  $4$ & $4$ & $-1$  & {\bf L} &           $1$  & $1$ & $1$ & $0$ & $0$ & $1$ & $1$ & $0$  \\ \hline
7& $M$ &  $2$ & $1$ & $\rho$  & {\bf T} &           $1$  & $-\rho$ & $2M\rho$ & $-\rho$ & $-1$ & $2M$ & $0$ & $0$  \\ \hline
8& $2$ &  $3$ & $3$ & $-2$  & {\bf L}   &     $1$  & $1$ & $1$ & $0$ & $0$ & $1$ & $1$ & $0$   \\ \hline
9& $3$ &  $2$ & $2$ & $-3$ & {\bf L} &            $1$  & $1$ & $1$ & $0$ & $0$ & $1$ & $1$ & $0$   \\ \hline
\hline
 \end{tabular}
\caption{All chiral spectra without massless charged free leptons that can be obtained for all $M$ and $N$ with $q_a \not=0$ and $q_b\not=0$. In item nr. 7 the value of $M$ is 1 or 2.}
\label{FreqTable}
\end{center}
\end{table}

Spectrum nr. 1   can be obtained from the Standard Model by interchanging  the r\^ole of {\bf U} and {\bf S}, and replacing color  by a mere multiplicity. 
It is built out of the combination of Higgs multiplets $3 {\cal H}(\frac12,1)+{\cal H}(\frac12,-\frac12)$.
The field {\bf U}
has no massless states, but its presence cancels the ``SU(1)" anomaly in the first factor.  
The corresponding string sector still exists, and starts at the first excited level.  
The low energy spectrum is just QED with charges proportional to $\pm1, \pm2$ and $\pm3$.
Note that here and in the following we divide all electromagnetic $U(1)$ charges in the low energy spectrum by their largest common denominator.

Spectrum nr. 2 can be written in terms of the Higgs multiplets
${\cal H}(1,-2)={\bf T}+{\bf Q}+{\bf L}+{\bf E}$,   ${\cal H}(\frac12,3)={\bf Q}+{\bf D}+{\bf E}$ and ${\cal H}(\frac12,1)={\bf L}+{\bf E}+{\bf N}$, where ${\bf N}$ is a singlet.
One family is equal to
${\cal H}(1,-2)+{\cal H}(\frac12,3)+{\cal H}(\frac12,1)$. So all fermions can indeed get a mass from the Higgs. The low energy spectrum has charges proportional to $\pm1$ and $\pm2$.

%

Spectrum nr. 3 has complete charge pairing for any solution of the anomaly cancellation conditions. Hence even after the pairing requirement we are left with a three parameter
family of spectra. Three independent combinations are shown in the table. All three can be written in terms of Higgs multiplets, but this requires adding some mirror pairs. We will
omit the details.
The low energy theory has charges proportional to $\pm1$ and $\pm2$. This is the only example we have found that
does not have automatic family repetition: different families can have a different structure.

All of these spectra have only a very limit number of possible charges, and no strong interactions to make
larger ones. So their anthropic prospects are bleak.


Spectrum nr. 4 
consists entirely of $SU(3)\times U(1)$ Higgs multiplets. The Higgs representation is ${\bf L}=(1,V^*,2)$. 
This solution is related to solution nr. 8 in the same way as nr 1. is related to the Standard Model.  Solution nr. 8 was already 
described above, and has a low energy spectrum (\ref{WrongSM}); it is like  the Standard Model, but with the color triplet Higgs. To obtain spectrum nr. 4 from spectrum nr. 8 one has to replace
the $SU(2)$ dimensions by a mere multiplicity, and let {\bf S} play the r\^ole of {\bf U}. The Higgs multiplets are essentially
the same as (\ref{SpecialHM}), with minor modifications:
\begin{equation}
(1,{\cal H}(\overline{V},-4))={\bf S}+{\bf E}+{\bf L}+{\bf N}; \ \ \ \ \ \ 2\times (1,{\cal H}(V,1))={\bf Q}+{\bf D}
\end{equation}
%
%
The low energy spectrum consists of $SU(2)$ doublets with charges $\pm 1$ and $0$, and singlets with charges $\pm 1, \pm 2$.
A family has the form
\begin{equation}
\label{NumberFour}
2(3,\tfrac13)+2(1,-1)+(1,2)+(\overline{3},\tfrac23)+(\overline{3},-\tfrac43) \ ,
\end{equation}
where the unbroken gauge group is $SU(3)\times U(1)$, and the particles are in the canonical order, \bf Q}, {\bf D}, {\bf S},
{\bf L}, {\bf E}. We have omited {\bf N}.
After symmetry breaking this becomes
\begin{equation}
\label{NumberFour}
(2,1)+(2,-1) + 2(2,0)  + 2(1,1)+2(1,-1)+(1,2)+(1,-2)
\end{equation}
To get any interesting atomic physics out of this spectrum one will have to assume that the $SU(2)$ interaction binds the
$SU(2)$ doublets into massive ``nuclei" 
with a non-trivial range of charges. Note that all $SU(2)$ bound states are bosons. This excludes anything like nuclear physics
in our universe, where the fermionic nature of protons and neutrons is crucial. 
Furthermore the $SU(2)$ interaction results from a Higgs mechanism
at a scale that must be low, in order to create a hierarchy. At this ``weak" scale, $SU(2)$ becomes part of a larger group $SU(3)$, and there
will be weak bosons that can convert  the $SU(2)$ doublets into $SU(2)$ singlets. Hence there will be interactions converting hadrons
into leptons; the details depend on the fundamental fermion masses.  Further analysis is needed to see if these flaws are fatal, and
for now we classify this case as ``acceptable".

%
%
%

Spectrum nr. 6 is also related to the Standard Model (nr. 5 is discussed below in section \ref{AsymTensorHiggs}).
Since $N=4$ the representation $A$ is an anti-vector of the subgroup $SU(3)$, and could pair off with a vector. 
This vector may come from either {\bf Q} or {{\bf X}, but not both. The
former requires $3Nq_b=-(N-1) q_a$, and $Q=E$. All fermions 
get a mass from a Higgs multiplet that consists of an entire family all by itself:
\begin{equation}
{\cal H}(V,3)={\bf D}+{\bf Q}+{\bf E}+{\bf L}+{\bf N};\ \ \ \  H=(V,-1)
\end{equation}
Hence all multiplicities must be equal to $Q$. The low energy spectrum is
\begin{equation}
\label{LikeSM}
Q \left[ (3,\tfrac23) + (1,1)+ (3,-\tfrac13) + {\rm c.c}\right]+Q(1,0)
\end{equation}
so this is just the low energy Standard Model with $Q$ families.

This corresponds to the $SU(5)$ GUT, broken to $SU(4)\times U(1)$, and using an $SU(4)$ vector as a Higgs boson. This is a solution
to our conditions, but it is also clear that it has major problems if we look more closely. The ``weak" interactions in this Universe couple the components of the broken $SU(4)$ multiplets to each other, and hence couple
quarks to leptons. Therefore there would be a catastrophic proton decay rate.

Indeed, the baryon violating operator is exactly one of the two one encounters in $SU(5)$, the one due to exchange of the vector boson usually called Y, with charge $\pm \frac13$.
In $SU(5)$ GUTs, this could for example give rise to the decay $p\rightarrow e^+ \pi_0$. Assuming that the Higgs system allows us to give all quarks and leptons exactly the same masses
as in our Universe, we can mimic nuclear and atomic physics in our Universe exactly. But then we would also know that proton decay is catastrophic, because by our assumptions the Yukawa
couplings are of order 1 ({\it i.e.} smaller than 1 by at most a few orders of magnitude, as in our Universe), and the Higgs scale then has to be around 100 GeV.  Of course one could consider changing
all the masses, and find another habitable range for totally different quark and lepton masses. One could even consider making the electron heavier than the proton (ignoring claims that the large
electron/proton mass ratio is anthropically important). But even then the process  $p+ e^- \rightarrow \gamma\gamma$ is always possible, and would be catastrophic as well.

This particular model satisfies all our constraints, and we label it therefore as ``acceptable". But with this very plausible additional argument it can be ruled out. It is quite clear why in our
Universe we find the Standard Model, and not its $SU(4)\times U(1)$ cousin described above.

Spectrum nr. 9 requires no further discussion: it is the Standard Model. The  alternatives to it have no low energy strong gauge group
at all, or only one or two $SU(2)$ factors. In addition to the models in the table there is the lepton-less series discussed in section \ref{QbIsZero}.  We already commented on the latter in that section. We will not attempt to dismiss the $SU(2)$-based models for lack of knowledge of strong $SU(2)$ interactions. But we think the results speak for themselves.


\subsubsection{Anti-symmetric Tensor Higgs}\label{AsymTensorHiggs}

Now let us discuss the anti-symmetric tensor Higgs for $M=1$.
If $H={\bf E}$ and $N$ is even, there is no $U(1)$ left over if the symmetry breaks to $Sp(N)$. If the symmetry breaking pattern is
as in Eqn. (\ref{AsymTensorHiggs}), the fields {\bf D} and {\bf S} must vanish, because they do not couple to the Higgs and are charged. For
$N=4$ the same is true for the anti-symmetric component of {\bf E}. From the anomaly conditions (\ref{Aneq}) we see that $D=S=E=0$ requires $M-N=4$, so that
$M=1, N=4$ is not a solution. If $N > 4$, and $D=S=0$ there are no charged singlets in the spectrum, so we cannot rule it out on the basis
of charged free leptons.  But it would be ruled out if we use condition 5b, because for $E\not=0$ the broken spectrum is non-chiral. 

Finally, consider the breaking of $SU(N)$ to $Sp(N-1)\times U(1)$ for $N$ odd. For $N \geq 5$, we must require $S=D=0$ for the same
reason as before. Furthermore $T$ contains a charged free lepton, and {\bf T} does not couple to $H={\bf E}$. Hence we must require $T=0$.
Substituting $S=D=T=0$ into the anomaly conditions we find that $M+N=4$, hence $N=3$, and  all other values are ruled out. Unlike the previous
paragraph, condition 5a is already sufficient here.

It is clear from the last paragraphs that $N=3$ is special. In this case the two possible breaking patterns give the same result, so we
consider only (\ref{Symplectic}). The fact that the anti-symmetric tensor $A$ of $SU(3)$ is an anti-vector gives
additional options for field-theoretic Higgs couplings. 
The Higgs {\bf E} can still not couple to {\bf T}, but it can couple to
{\bf S} and {\bf D} for special values of $q_a$ and $q_b$. Rather than working out all possible Higgs couplings we
will determine all possible values of $q_a$ and $q_b$ that allow pairings of the singlets in (\ref{Symplectic}).
These singlets and their charges are (with $q_a=1$ and $\tilde q_b=Nq_b \equiv \rho$)
\begin{eqnarray*}
Q &\rightarrow& 1+\rho \\
D &\rightarrow& -1 \\
S &\rightarrow& 2 \\
X &\rightarrow& 1-\rho \\
L &\rightarrow& -\rho 
\end{eqnarray*}
Because of the symmetry of the anomaly conditions and because $q_b \not =0$ we may require that $\rho < 0$.
Now consider all ten possible pairings, each with both signs. We find that there are four
values of $\rho < 0$ that allow special charge pairings namely,  $\rho=-\frac12, -1,-2$ or $-3$. For each
value of $\rho$ the multiplicities must be matched correspondingly.  
For the four solutions these pairing conditions (including $T=0$) are  respectively
\begin{eqnarray*}
T=D=S=X=0;\ \ \ Q+L=0\\
T=0;\ \ \ D-L=0; \ \ \  S+X=0\\
T=0; \ \ \ Q+D=0;  \ \ \  S+L=0;\ \ \ X=0\\
T=0;\ \ \  Q-S=0;\ \ \ D=X=L=0
\end{eqnarray*}
In the first, third and fourth case this implies that all multiplicities must vanish, but there is a solution for
the second pairing:
\begin{equation}
U=Q;\ \ L=D=-2X;\ \ S=-X;\ \ E=X+Q;
\end{equation}
This is a two parameter solution. Note however that for $\rho=-1$ we have $q_b=-\frac13$.  Then the fields {\bf Q} and 
{\bf E}  are $(1,3,\tfrac23)$ and $(1,\overline{3},-\tfrac23)$ and hence can pair off without Higgs symmetry breaking.
This implies that we can set $Q=0$.  Then we get  a one-parameter solution listed as item nr. 5 in the table.
The unbroken $SU(3)\times U(1)$ spectrum is
\begin{equation}
2(1,1)+(1,-2)+({\overline 3},\tfrac43)+2(3,-\tfrac13)+({\overline 3},-\tfrac23) \ ,
\end{equation}
where the fields are {\bf D}, {\bf S}, {\bf X}, {\bf L}, {\bf E}, respectively.
The Higgs breaks this to
\begin{equation}
\left[2(1,1)+(1,2)+(2,1) + \hbox{c.c.}\right]+2(2,0)+(1,0) \ .
\end{equation}
It turns out that this case is almost identical to item nr. 4 in the table, although it
is realized in a rather different way. Note that the cancellation of the stringy ``SU(1)" anomaly
is distinct, but the resulting chiral $SU(3)\times U(1)$ spectrum is identical, up to charge conjugation. 
Because of the different charge coefficients $q_a$ and $q_b$ in the two cases,
the Higgses {\bf L} in spectrum nr. 4 and {\bf E} in spectrum nr. 5 are in the same $SU(3)\times U(1)$ representation
and break the spectrum in the same way. The comments given above for spectrum nr. 4 apply here as well.

%
%
%

\section{Phenomenology of  $U(3)\times U(2)$ models.}\label{GutlessGut}

The high energy gauge group we identified as a potentially superior alternative to $SU(5)$ GUTs is not unknown. It
 is often mentioned in text books as $S(U(3)\times U(2))$, a group-theoretic description of charge quantization.  
It is not usually taken seriously as a fundamental group, because it looks just like a first step towards $SU(5)$, and with a strong prejudice towards symmetries
it seemed obvious that this was its reason for existence. But in the context of brane models this possibility starts looking a lot more viable. 
The unitary gauge groups provided by brane models are $U(N)$, with a unitary phase factor built in.
In brane models, there is
always a subclass where a stack of five-branes is split into a stack of three and a stack of two. But there is not really any fundamental reason why those stacks have
to be on top of each other at {\it any} scale. Furthermore, there are numerous examples where the two stacks are made of unrelated branes, that cannot
even be put on top of each other. Intuitively, one  would expect separate three-stacks and two-stacks to be more generic  than a five-stack. 

We do not intend to propose a concrete realization of such a model here. Numerous examples of spectra in this general class have been
obtained in a scan of Gepner models  \cite{Anastasopoulos:2006da}. They belong to the class identified as ``$x=0$" in that paper. {Moreover, brane realizations
of $U(3)\times U(2)$ models have been described  before in \cite{Ibanez:1998rf} (see in particular section 3 of that paper).}
So there is no doubt that such models can indeed be realized.
Here we will just make a few remarks about their phenomenological features.

Since this class contains $SU(5)$ models as a subclass, $U(3)\times U(2)$ models are phenomenologically just as viable. They give more freedom than
$SU(5)$ at the price of being less predictive. But in order to make predictions based on the $SU(5)$ subset, one has
to come up with a convincing reason why we should find our Universe within that subset. We have shown that neither
charge quantization nor the structure of a family offers such a reason. Brane model building does not offer such a reason either. Perhaps
F-theory does, but such a conclusion requires a much better understanding of the various options available in non-perturbative string theory in general.

Note that our arguments can also be used within the context of traditional $SU(5)$ model building to rule out the symmetric tensor
representation $({\bf 15})$. Furthermore one can invoke similar anthropic arguments to solve the doublet-triplet splitting problem in that
context as well as motivate the GUT Higgs breaking breaking to $SU(3)\times SU(2)\times U(1)$ rather than to $SU(4)\times U(1)$
(provided models 6 and 8 in our table can be convincingly ruled out anthropically). But in {\it none} of these cases having a full $SU(5)$ 
 offers any advantages; the $U(3)\times U(2)$ model is superior in all respects, apart from aesthetics. But aesthetics may well be the wrong
 guiding principle in the post-SM era.

\subsection{Coupling constants}

In $U(M)\times U(N)$ brane models there are only two fundamental gauge couplings, whose size is determined by
the dilaton couplings to the branes. Their relative size is determined by the volume of cycles of the compactification 
manifold on which the branes are wrapped. In any case, this gives two free coupling constants at the string scale, which 
we will denote $g_a$ and $g_b$. There are $M^2$ $U(M)$ gauge bosons  $A_{\mu}^i$ and $N^2$ $U(N)$ gauge bosons  $B_{\mu}^j$, and
we will denote the  $U(1)$ generator of $U(M)$ or $U(N)$ by an index $i=0$ or $j=0$.
We denote the $U(M)$ and $U(N)$ generators as $T$ and $W$, respectively. The gauge interaction with the fermions $\psi$  takes the form
\begin{equation}
g_a \bar \psi \gamma^{\mu} \sum_{i=1}^{M^2-1} A_{\mu}^i T^i \psi+g_b \bar \psi \gamma^{\mu} \sum_{j=1}^{N^2-1} B_{\mu}^jW^j \psi
+g_a \bar \psi \gamma^{\mu} A_{\mu}^0  T^0\psi
+g_b \bar \psi \gamma^{\mu} B_{\mu}^0 W^0 \psi
\end{equation}
The first two terms  yield the strong and weak interactions. We choose the standard normalization (for all labels, including 0)
\begin{equation}
\label{GenNorm}
{\rm Tr}\ T^i T^j =\frac12 \delta^{ij} \  , \ \ \ \  \ {\rm Tr}\ W^i W^j =\frac12 \delta^{ij} \ \ \ \  
\end{equation}
which guarantees that all gauge bosons have canonically normalized kinetic terms.
From the first two terms we read of the Standard Model $SU(3)$ and $SU(2)$ gauge interactions in their
standard form. Hence the Standard Model couplings are $g_3=g_a$ and $g_2=g_b$. The generators of the $U(1)$ components are related in the following way to the brane charges
\begin{equation}
T^0=\frac{1}{\sqrt{2M}}Q_a\ , \ \ \ W^0=\frac{1}{\sqrt{2N}}Q_b
\end{equation}
The gauge boson $A^Y_{\mu}$ of the Standard Model $U(1)$ factor $Y$ is an orthogonal rotation of $A^0_{\mu}$  and  $B^0_{\mu}$:
%
%
%
%
\begin{eqnarray*}
A^0_{\mu}={\rm sin}(\theta) A^Y_{\mu}+{\rm cos}(\theta) R_{\mu}\\
B^0_{\mu}={\rm cos}(\theta) A^Y_{\mu}-{\rm sin}(\theta) R_{\mu}
\end{eqnarray*}
where $R$ is the orthogonal component that gets a mass from the Green-Schwarz mechanism in string theory.
Hence the massless components couple as follows to matter
\begin{equation}
g_a \bar \psi \gamma^{\mu} {\rm sin}(\theta) A^Y_{\mu}  T^0\psi
+g_b \bar \psi \gamma^{\mu} {\rm cos}(\theta)A^Y_{\mu} W^0 \psi
\end{equation}
%
%
Comparing the coupling of $A_{\mu}^Y$ with the Standard Model we get
\begin{equation}
g_a {\rm sin}(\theta)\frac{1}{\sqrt{2M}}Q_a+g_b {\rm cos}(\theta)\frac{1}{\sqrt{2M}}Q_b=g_Y (q_a Q_a + q_b Q_b)\ ,
\end{equation}
where $q_a=-\tfrac13$ and $q_b=\tfrac12$ is the correct normalization to obtain the usual Standard Model $Y$ charges, so that
the quark doublet is in the representation $(3,2,\frac16)$. The ratio of the coefficients of $Q_a$ and  $Q_b$ determines 
${\rm sin}(\theta)$ as
\begin{equation}
{\rm sin}^2(\theta)=\frac{M g_b^2 q_a^2}{N g_a^2 q_b^2+Mg_b^2 q_a^2}
\end{equation}
and then we get the following result for $g_Y$
\begin{equation}
g_Y^2 =g_1^2 {\rm sin}^2(\theta)=\frac12 \frac{g_a^2 g_b^2 }{N g_a^2 q_b^2+Mg_b^2 q_a^2}
\end{equation}
For $N=2, M=3, q_a=-\tfrac13, q_b=\tfrac12$ this yields
\begin{equation}
\label{CoupleRel}
g_Y^2 =g_a^2 {\rm sin}^2(\theta)=\frac{3 g_a^2 g_b^2 }{3 g_a^2 +2 g_b^2}
\end{equation}
For $SU(5)$ ($g_a=g_b\equiv g$) this yields the familiar result $g_Y=\sqrt{\frac{3}{5}}g$.
The relation (\ref{CoupleRel}) can be written as
\begin{equation}
\label{CoupleRelTwo}
\frac{1}{\alpha_Y} = \frac{2}{3} \frac{1}{\alpha_s} + \frac{1}{\alpha_w}
\end{equation}
{
This agrees with \cite{Ibanez:1998rf}.
Precisely the same relation was found in \cite{Blumenhagen:2006ci} for a class of Pati-Salam models. In this
class there is a relation between the three gauge couplings of $SU(4)\times SU(2)_L \times SU(2)_R$   if the two $SU(2)$ factors have a related
brane origin. More recently, the same relation was found in a class of $U(5)$ F-theory models with hypercharge flux breaking \cite{Ibanez:2012zg}. }

Extrapolating the measured coupling constants to higher energies from their values at 100 GeV ($g_1=.357, g_2=.652, g_3=1.212$)
 we find that relation (\ref{CoupleRelTwo}) is satisfied at a scale
$M_{\hbox{non-susy}}=10^{13.76}$ GeV}, with
\begin{equation}
\   g_1=0.5511,\  g_3=g_a=0.570\ \hbox{and }\  g_2=g_b=0.5391
\end{equation}
where we used the non-supersymmetric $\beta$-function coefficients. With supersymmetric $\beta$-functions and a susy breaking scale
at 1 TeV we find $M_{\hbox{susy}}=10^{16.15}$ GeV, with
\begin{equation}
  g_1=0.699,\  g_3=g_a=0.696\ \hbox{and }\  g_2=g_b=0.702
\end{equation}
In the supersymmetric case the scale where (\ref{CoupleRelTwo}) holds is of course the usual susy-GUT scale, and there is an obvious candidate for
the physics associated with
that scale: GUT unification. In the non-supersymmetric case there is no unification into a larger gauge group, and it is less obvious
what happens. 

The natural guess is that at $10^{13.76}\ \hbox{GeV}=5.75 \times 10^{13}\  \hbox{GeV}$ we reach the string scale, 
and that the gauge groups $U(M)$ and $U(N)$ are
described by a Dirac-Born-Infeld action at that scale. But 
there are other possibilities if one allows dimensions to decompactify at different scales, and the result also depends on
the dimension of the branes on which the unitary groups live. We will not pursue this point further in this paper. The aforementioned
scale just gives a rough indication of the location of ``new physics" in this class of models.

%
%

\subsection{Yukawa couplings}

In this paper we have only considered field-theoretic selection rules for Yukawa couplings. Perturbative brane models of this
type have a well-known problem with Yukawa couplings for the anti-symmetric tensor fields. The problem is that these couplings
 are forbidden by the two $U(1)$ factors of $U(3)$ and $U(2)$. Both of these $U(1)$'s are broken by the Green-Schwarz mechanism,
 leaving only the Standard Model $U(1)$ that does not forbid these couplings. However, the two $U(1)$'s remain as global
 symmetries of the perturbative spectrum, that still do not allow the desired couplings. 
 
 The solution of this problem requires non-perturbative physics. Indeed, it has been shown that brane instantons can generate these
 couplings \cite{Blumenhagen:2007zk}. These instantons do not have to be solutions of one of the gauge groups, but they can be stringy instantons associated with
 a brane that does not contribute to the gauge group, which have been proposed as solutions
 to various other problems \cite{Florea:2006si,Blumenhagen:2006xt,Ibanez:2006da,Ibanez:2007rs,Argurio:2007vqa}. In GUT models,
 generating Yukawa couplings is problematic, because often the same instantons generate baryon number violating interactions \cite{Kiritsis:2009sf}.
 This problem can be avoided by adding extra branes to the configuration, with somewhat ad-hoc assignments of $U(1)$ charges 
\cite{Anastasopoulos:2010hu,Anastasopoulos:2011zz}. But without SU(5) unification, this may not be  needed. See also section
\ref{ProtonDecay} below.
 
 Since one of the perturbatively forbidden Yukawa couplings is the one of the
 top quark, the non-perturbative effect cannot be small, therefore one has to find a genuinely non-perturbative description of such a model.
 In the case of $SU(5)$ susy-GUTs, F-theory has been proposed as a solution \cite{Beasley:2008dc,Heckman:2008qa,Cecotti:2009zf}. There is no reason to believe that something similar could
 not work for non-supersymmetric $S(U(3)\times U(2))$ models, although it might be hard to construct explicit models without the 
 powerful tools of supersymmetry.

\subsection{Neutrinos}

There are no candidates for right-handed neutrinos in the $U(3)\times U(2)$ spectrum (\ref{BraneReps}), but this is only because we did not take singlets into account. Note that the vector representations {\bf L} and {\bf D} are obtained from open strings with one end attached to
an object outside the $U(M)\times U(N)$ configuration. Then there will also exist open strings with both ends on a neutral object. If they
are fermions, all that is required is a Dirac coupling to {\bf L} and $H$. Nothing forbids such a coupling. 
Three independent linear combinations coupling to the three species of {\bf L} is all that is needed.
With a sufficiently large number
of singlets,  these are likely to exist. The right-handed neutrinos would have Majorana masses of order $10^{14}$ GeV, the string scale. 
This would be roughly the correct order of magnitude
for Yukawa couplings comparable in size to those of the charged quarks and leptons (which are spread over a large range anyway). 


\subsection{Magnetic monopoles}
There is a folk theorem stating that charge quantization goes hand-in-hand with monopoles satisfying the corresponding Dirac
quantization condition, $eg=2\pi m, m \in \mathbb{Z}$ \cite{DiracMonopole}. 
In $SU(5)$ field theoretic models, such monopoles can indeed be constructed as solutions
to the classical field equations. In the subclass of $S(U(3)\times U(2))$ models that correspond to continuously deformed
$SU(5)$ models, they will then also exist. However, in the general class of $S(U(3)\times U(2))$ models we only know that the perturbative
states satisfy the observed charge quantization. In principle, in some cases there could exist non-perturbative solutions with a charge
that is not an integer multiple of the electron charge.
If the folk theorem holds (though this has been questioned recently 
\cite{Hellerman:2013vxa,Hellerman:2013mpa})
there are then two possibilities for the full class of models: either there are non-perturbative
states corresponding to monopoles with the minimal Dirac magnetic charge, or there are non-pertubative states with unobserved
fractional electric charges ({\it i.e.} color singlets with charges that are not an integer multiple of the electron charge). The latter outcome would be disappointing, but not fatal, since 
 these non-perturbative states could well be very massive.

\subsection{Proton decay}\label{ProtonDecay}

Without an explicit $SU(5)$ there is no need to worry about many of the issues that plague $SU(5)$ GUTs, such as the doublet-triplet splitting problem or 
alignment of the two Higgs systems so that color is not broken.  However, it is not necessarily true that proton decay is avoided. Note that
baryon and lepton number are not symmetries of these models. The two-stack model has two $U(1)$'s, $Q_a$ and $Q_b$, that are in general anomalous and broken
by a Green-Schwarz mechanism. These symmetries remain as exact global symmetries in the perturbative theory and may be broken by instantons.
But these symmetries are {\it not} baryon number and lepton number. This is due to the use of rank two tensors in the spectrum; indeed, in standard models of the Madrid type, there are perturbative baryon and lepton number symmetries. 

In addition to the Standard Model fermions and the Higgs boson, the $U(M)\times U(N)$ brane model allows many
other particles. 
They can be absent in the massless spectrum, but it is not reasonable to assume they are absent altogether. 
For example, there can be massive vector bosons in the representation ${\bf X}$. It is easy to check that a vector boson in that representation can have the usual $SU(5)$ couplings with
quarks and leptons that lead to proton decay, without violating $Q_a$ and $Q_b$. This is inevitable, since this class of models contains broken $SU(5)$ as a special case. 
However, without having an explicit $SU(5)$, the presence of such particles in the spectrum is not automatic. Scalars in any of the
allowed 
representations are potentially equally dangerous. Indeed, if there are light scalars in the representations {\bf Q}, {\bf U}, {\bf D}, 
{\bf E} or {\bf L} it is clear that one can write down dimension 4 baryon or lepton number violating operators like the infamous ones of the MSSM. Without low energy
supersymmetry there is no need for them to be light, and under our assumptions no scalars are light unless  they are needed.

However, we still have to worry about any of these particles in the massive spectrum. It is clear that generically, even if they
are not light but have masses at the string scale they are still problematic, because  the
estimated string scale is about $5.7\times 10^{13}$ GeV. This creates a potential conflict with existing limits on dimension 6 proton decay,
which would require a scale about two orders of magnitude larger, assuming all other factors are of order 1. 
In comparison to the problems of susy-GUTs with proton decay by dimension 4 and  dimension 5 operators, this does not look like
a huge problem, which may be overcome in specific models. 

Although the $U(3) \times U(2)$ class we consider is
less predictive than $SU(5)$ GUT phenomenology, we are
 limited by some self-imposed landscape naturalness restrictions. 
Thus if one takes the point of view that the Standard Model is part
of a huge landscape,
as we do, one should not need tricks, like additional Higgses or discrete symmetries, that are deviously constructed just to hide generic properties, unless that
property is fatal for the existence of life. For example, one has to distinguish proton decay that can be observed
and proton decay that destroys the observer. In the latter case some new physics may be postulated, but not in the former.
Proton decay by dimension 4 operators in supersymmetric theories 
is anthropically catastrophic, and hence it is legitimate to require discrete symmetries to forbid it. Here ``require" means to select from
the available models just those that have a chance of being observed.
On the other hand, proton decay by dimension 5 operators is merely observable, but not fatal, and should be absent without excessive 
model building efforts.

Proton decay in non-supersymmetric $U(3) \times U(2)$ models with the Higgs as the only light scalar is in any case too small to be anthropically
relevant, because at worst it goes via dimension 6 operators. Then the  relevant landscape naturalness question  is
in which fraction of the  models that allow the existence of observers,  observable (but not fatal) proton decay is avoided. If it turns out that this is true only in a very small fraction, we would
consider this class as falsified.

\subsubsection{Supersymmetry?}

Our conclusions on charge quantization and family structure remain valid
in the presence of supersymmetry. Of course low energy supersymmetry would have the advantage of raising the string scale
and thereby reducing proton decay by dimension 6 operators.
But this is the worst way to prevent 
dimension 6 proton decay, because
it introduces dimension 4 and 5 operators that are even worse.

However, it has been observed in \cite{ArkaniHamed:2004fb,Giudice:2004tc,ArkaniHamed:2004yi} that one can keep some of the good features of supersymmetry
while removing the worst part, an idea that might well have been called ``susy without guts", but is actually known as 
``split supersymmetry". {One may also simply raise the entire supersymmetry breaking scale. For a well-motivated scenario with intermediate scale supersymmetry and
some phenomenological features similar to the ones discussed here, see \cite{Ibanez:2012zg}.}
Even without  a supersymmetric theory, {the main  idea of split supersymmetry } can be implemented by assuming that in addition to the Standard Model there are some fermions in the adjoint representation
of the gauge group, with the same properties as gauginos, but not necessarily related to  supersymmetry. 

In the introduction we argued that perhaps we observe the current Standard Model  because having several light
fermions is statistically too costly, and  a single Higgs is more economic. This argument would seem to go against the
existence of light gauge fermions simply to increase the life of the proton beyond observational bounds. But a gauge fermion
is a different object than any of the matter fermions. It is imaginable that it has an entirely different mass distribution, perhaps
even a scale invariant one, cut off at the lower end by an anthropic dark matter constraint. Particles with a scale invariant mass
distribution can be light at no cost.
Hence some of these particles might end up somewhere in the light spectrum, at a scale not
(necessarily) related to the weak scale. They may also have different multiplicities and a different mass distribution than
expected on the basis of supersymmetry. 

It is easy to construct examples where the mass scale is raised, but the
three gauge couplings do not pass through the same point. 
The discovery of gaugino-like particles that raise the string scale, but fail to make the couplings convergence, would mean the end of susy-GUTs, and would leave GUTs without guts as a viable solution to the charge quantization problem. On the other hand, if
new matter (supersymmetric or not) is found at the LHC or elsewhere that makes the three gauge coupling converge this would still be consistent
with $S(U(3)\times U(2))$ models, but undoubtedly this would be generally seen as evidence for $SU(5)$ unification. However,
it is important to keep in mind that this would be based purely on aesthetics, and that any scientific argument in favor of unification
based on the Standard Model spectrum or charge quantization is invalid. The single example we discussed in this paper suffices to
demonstrate that.

Can we ever determine which of these possibilities is realized in Nature?
Perhaps the most attractive scenario is that proton decay is finally
discovered after all, and can help disentangle the various scenarios.

\section{Conclusions}\label{Conclusions}

Our main conclusion is that in a class of models that includes $SU(5)$ GUTs, charge quantization can be 
understood without it, and holds in a much wider set of models. Given the Standard Model gauge group, the 
quark and lepton structure of a family is also determined
uniquely, and hence one automatically gets repetition of families with identical structure. No GUT group structure is  
needed for any of this.

We replace the top-down GUT idea by a bottom-up requirement that is
 the main selling point of the Higgs mechanism: that it gives mass to all fundamental fermions. 
This turns out to be so powerful that it selects the Standard Model almost uniquely 
in the set of $U(M)\times U(N)$ brane models, including its gauge group and the Higgs boson, assuming
a strong interaction group $SU(M)$. The only remaining alternatives
are some purely electromagnetic theories with a too limited set of charges,
a class without leptons and probably no baryogenesis, and some theories with strong interactions based on $SU(2)$, 
whose spectrum we cannot reliably determine, but which have potentially fatal flaws.

Are we just marvelling at our own beautiful blue planet with its atmosphere full of oxygen, while other creatures
might be marvelling at their ugly brown planet with its cyanide abundance? Did we walk into the anthropocentric trap?
We believe this is not the case, because in this context the Standard Model can be determined by objective criteria.
Any dedicated graduate student with some knowledge of string theory and brane models, but
without any knowledge of Particle Physics or the Standard Model would eventually
stumble on it by merely investigating the Higgs mechanism in two-stack brane models.

Although the condition that the Higgs gives mass to all fundamental fermions may seem an ad-hoc
phenomenological requirement, a 
slightly weaker form  is plausibly needed for anthropic reasons: a massless photon and no massless charged leptons. 
This weaker form is harder to solve in
general, especially if one allows for chiral symmetry breaking by non-abelian groups. But trying this
in cases under sufficient control produces no additional solutions.

To our knowledge this is the most convincing determination of the Standard Model structure in any context.
What makes it so convincing in comparison to other approaches that can claim some success, such as GUTs or
noncommutative geometry \cite{Chamseddine:2013rta}, is  the fact that one of our conditions, the existence of observers,
is unquestionably
a necessary condition, unlike symmetry or geometry. Although anthropic arguments have been dismissed as tautological by some
and anti-scientific by others, this is an impressive example of their power in comparison to the traditional methods based on
abstract mathematical concepts.  

We have to admit that a  mystery remains: why does the anthropic solution coincide with a broken $SU(5)$ representation? 
Perhaps this should be seen as one of those coincidences that occur in the theory of Lie algebras, where there exist 
isomorphisms between low rank Dynkin diagrams, such as $C_2 \sim B_2$. The Standard Model is not a unique solution
to our criteria, and if we were to allow more branes we know that an infinite series of potential alternatives would appear with
gauge group $SU(M)\times SU(2)\times U(1)$. Perhaps as the first entry in that series the Standard Model just happens to have
an atypical realization as a broken GUT. 

The two-stack brane configuration we started with is our strongest assumption. We hope to relax this in future work.
It should be possible to consider string configurations with more branes, including ones that contain the Madrid model \cite{Ibanez:2001nd}. 
The assumption that there is a strong $SU(M)$ gauge group may also be relaxed in order to investigate if anthropic universes can exist without it.
One may also consider
bi-fundamental Higgses that break two unitary factors. 
It may also be possible to strengthen the constraints by incorporating 
other potentially anthropic requirements. For example, we have almost completely ignored the effect of the weak interactions.

The requirement that all fermionic matter gets a mass from a single Higgs can also be applied to heterotic strings, but it will not solve the problem
of fractional charges in that context. There will in any case be massive fractionally charge particles \cite{Schellekens:1989qb},
but there may also be chiral ones that get their mass from the Higgs. There exist fractionally charged Higgs multiplets, and
only a miraculous outcome of modular invariance could prevent these from being absent in all cases.

It is ironic that there are two main areas in string theory where Grand Unification has been extensively discussed, and
in both cases it does not deliver its promise of explaining charge quantization: in brane models it is not needed, and
in heterotic strings it does not work.

\vskip 2.truecm
\noindent
{\bf Acknowledgments:}
\vskip .2in
\noindent
{It is a pleasure to thank Jim Halverson and Luis Iba\~nez for thoughtful comments on the first version of this manuscript.}
This work has been partially 
supported by funding of the Spanish Ministerio de Econom\'\i a y Competitividad, Research Project
FIS2012-38816, and by the Project CONSOLIDER-INGENIO 2010, Programme CPAN
(CSD2007-00042).

\vskip .5in

\bibliography{Landscape}{}

\bibliographystyle{ieeetr}

\end{document}